\documentclass[showpacs,twocolumn,amsmath,amssymb,pre,aps]{revtex4}
\usepackage{amsmath}
\usepackage{amsthm} 
\usepackage{amssymb}
\usepackage{epsfig}
\usepackage{graphicx}% Include figure files
\usepackage{bm}% bold math
\usepackage{mathrsfs}
\usepackage{float}

\begin{document}

\title{ 
Controlling the transport of electrons on superfluid $^{4}$He in symmetric and asymmetric FET-like structures 
}
\draft

\author{A.A. Vasylenko and V.R. Misko }
\affiliation{
{Department of Physics, University of Antwerp, Groenenborgerlaan 171, B-2020 Antwerpen, Belgium }
}

\date{\today} 

\begin{abstract}
When floating on a two-dimensional surface of superfluid $^{4} $He, electrons arrange themselves in two-dimensional crystalline structure known as Wigner crystal. 
In channels, the boundaries interfere the crystalline order and in case of very narrow channels one observes a quasi-one-dimensional (quasi-1D) Wigner crystal formed by just a few rows of electrons and, ultimately, one row, i.e., in the ``quantum wire'' regime. 
Recently, the ``quantum wire'' regime was accessed experimentally [D.~Rees {\it et al.}~\cite{Rees-Totsuji-2012}] resulting in unusual transport phenomena such as, e.g., oscillations in the electron conductance. 
Using molecular dynamics simulations, we study the nonlinear transport of electrons in channels with various types of constrictions: single and multiple symmetric and asymmetric geometrical constrictions with varying width and length, and saddle-point-type potentials with varying gate voltage. 
In particular, we analyze the average particle velocity of the particles versus the driving force or the gate voltage. 
We have revealed a significant difference in the dynamics for long and short constrictions: The oscillations of the average velocity of the particles for the systems with short constrictions exhibit a clear correlation with the transitions 
between the states with different numbers of rows of particles; on the other hand, for the systems with longer constrictions these oscillations are suppressed. 
The obtained results are in agreement with the experimental observations by D.~Rees {\it et al.}~\cite{Rees-Totsuji-2012}. 
We proposed a FET-like structure that consists of a channel with asymmetric constrictions. 
We show that applying a transverse bias results either in increase of the average particle velocity or in its suppression thus allowing a flexible control tool over the electron transport. 
Our results bring important insights into the dynamics of electrons floating on the surface of superfluid $^{4}$He in channels with constrictions and allow an effective control over the electron transport. 
\end{abstract}
\pacs
{
73.23.-b,	%Electronic transport in mesoscopic systems
71.10.-w,	%Theories and models of many-electron systems
52.65.Yy,	%Molecular dynamics methods
85.30.Hi	%Surface barrier, boundary, and point contact devices
%
%73.20.Qt	Electron solids
%73.23.-b	Electronic transport in mesoscopic systems
%85.30.Hi	Surface barrier, boundary, and point contact devices
%45.50.Jf	Few- and many-body systems
%
%71.10.-w	Theories and models of many-electron systems
%52.65.Yy	Molecular dynamics methods
}
\maketitle

%Keywords{Wigner crystal; Electron transport; Point contact}

\section{Introduction}

Investigation of the transport of small particles through narrow constrictions is of essential importance in many physical systems. 
The transport properties of tiny particles have been studied with use of both experimental and simulation methods (see, e.g., 
\cite{Andrei97,2DCLbook,Rees-Kono-2010,Piacente2004,Piacente2005}). 
An ideal model system for studying strongly interacting electrons is a system of electrons floating on the surface of superfluid $^{4}$He \cite{Rees-Kono-2010,Araki-Hayakawa-2012}. 
The Wigner solid, or Wigner crystal, was predicted theoretically by Wigner in 1934 \cite{Wigner34}. 
In 1970s, the interest in the Wigner crystal as a real physical object grew significantly due to the new developments and discoveries in the studies of two-dimensional low-density electron systems in semiconductors and on a surface of liquid helium~\cite{Crandall71}. 
In 1979 the Wigner solid was first observed experimentally in a two-dimensional electron system on the surface of liquid helium~\cite{Iye80}. 

A system of electrons on a surface of liquid helium provides an opportunity for investigating concerted effects between electrons and a soft interface. 
A number of distinct phenomena associated with the interaction with a free surface of liquid $^{4}$He have been observed, in particular, in an electron crystal or the Wigner solid phase, where electrons are self-trapped in a commensurate surface deformation called the dimple lattice. 
On a flat surface of liquid $^{4}$He, the Wigner solid moves as a whole keeping the hexagonal lattice under a driving force parallel to the surface. 
The electron motion on liquid helium is associated with surface excitations, or ripplons (see, e.g., \cite{Tempere}). 
When travelling faster than the ripplon phase velocity, as in the case of the Cherenkov radiation, an electron radiates surface waves and the ripplons emitted by different electrons interfere constructively if the wave number of the ripplons equals the reciprocal lattice vector of the Wigner solid (the Bragg condition). 
This resonant Bragg-Cherenkov emission of ripplons gives rise to the limitation of the electron velocity, which was first observed in \cite{Kristensen} and analyzed in \cite{Dykman}. 
Another intriguing nonlinear phenomenon, a sharp rise in mobility at a much higher excitation, was found in \cite{Shirahama}. 
This effect was attributed to decoupling of the Wigner solid from the dimple lattice, and the observed features were qualitatively understood by decoupling from a rigid dimple lattice. 
Later a simple hydrodynamic model was proposed \cite{Vinen} based on the assumption that decoupling occurs from the dimple lattice that deepens due to the Bragg-Cherenkov scattering thus bridging the two above-mentioned phenomena.

%At low temperatures, a one-component plasma undergoes a phase %transition and forms a Wigner crystal both in two (2D) and 
%three dimensions (3D) \cite{Grimes}. 
For a finite 2D Wigner model of particles with $ln(1/r)$ interacting potential, the equation of state and some of the ground state configurations were reported in \cite{Calion}. 
Within a more realistic model using a Coulomb potential, it was shown in \cite{Bedanov} 
that circularly confined electrons arrange themselves in ring configurations 
(Wigner molecules, or ``Wigner islands''), which were recently
visualized in experiment \cite{Rousseau}. 
Using Monte Carlo simulations, the structural, dynamical properties and melting of a quasi-one-dimensional system of charged particles, interacting through a screened Coulomb potential were studied in \cite{Piacente2004, Piacente2005}. 

Recent advances in microfabrication technology have allowed the study of the Wigner solid in confined geometries using devices such as microchannel arrays 
\cite{Glasson,Ikegami}, 
single-electron traps \cite{Papageorgiou}, 
field-effect transistors (FET) \cite{Klier} 
and charge-coupled devices \cite{Sabouret}. 
However, a ``quantum wire'' regime when the effective width of a conductive channel is less than the thermal wavelength of the electrons has been only accessed recently, in experiments \cite{Rees-Kono-2010} where the transport properties of electrons were measured in a microchannel, 
with the confinement potential controlled on the scale of the inter-electron separation ($\approx$ 0.5~$\mu$m). 

Here, we focus on the electron transport in a microchannel with a width of the order of the inter-electron separation~\cite{Rees-Kono-2010}. 
We employ molecular dynamics simulations to study an externally driven Wigner solid 
(similarly to the system of charged colloidal particles driven in a quasi-1D channel~\cite{tkachenkoQ1D} or diffusing in a circular channel~\cite{tkachenkoSFD}). 

To understand the role of Wigner solid melting in the electron transport through a small microchannel, we refer to the results of the experiments~\cite{Rees-Kono-2010,Rees-Totsuji-2012}, where the current was found to be dependent on a tunable potential barrier formed in the channel by a split-gate electrode beneath the helium surface, as well as the intrinsic resistance of the electron system. 
The latter effect is more pronounced for small or zero gate voltage when the threshold in the transport current may be induced by the melting of the 2D Wigner solid in the microchannel. 
This effect is investigated in detail for different ratios of the inter-electron spacing to the channel width and varying temperatures, and the \textit{IV}-curves are calculated, without and in the presence of a gate voltage. 

Next, we propose an experimental set-up that includes a set of asymmetric constrictions along the channel. 
This structure is shown to function as a FET-like device: the electron current can be easily controlled in this device by applying a transverse bias. 
The electron flow induced by the external driving, can be directed either toward the side of the channel without constriction (and thus flow freely) or toward the constrictions where the particle motion is suppressed. 
In both cases, the electron transport is characterizied by striking oscillations in the average velocity of the particles. 

The paper is organized as follows. 
In Sec.~II, we introduce the model system. 
Sec.~III is devoted to the study of symmetric channels with various types of constrictions and the relevance of the simulation results to the experiments with electrons in the ``quantum wire'' regim. 
In Sec.~IV, we introduce an asymmetric FET-like structure, as a proposal for an experimental set-up, and investigate the trasport properties of this structure. 
Finally, in Sec.~V, we summarize our findings.

\begin{figure}[t!]
\centerline{\epsfig{figure=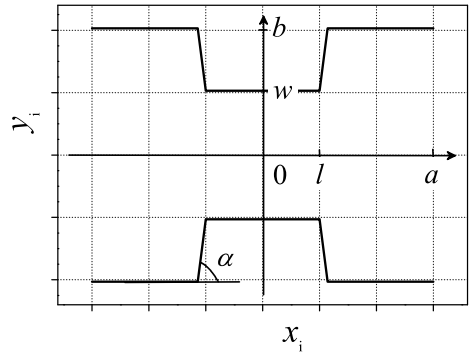,width=8cm}}
\vspace*{-0.5cm}
\caption{
Sketch of the simplest system: a channel with one constriction. 
}
\label{ChannelSketch}
\end{figure}

\section{Model}

We study the dynamics of electrons floating on surface of superfluid $^{4}$He driven by an external force in a channel with a constriction (see Fig.~\ref{ChannelSketch}). 
The equations of motion of the electrons in the $xy$-plane are as follows: 
\begin{eqnarray}
m\ddot{x}_{i}&=&-\eta \dot{x}_i+\frac{q^{2}}{\epsilon}\sum_{i,j} F_{x,ij}+F_{x,i}^{T}+F_{dr} \nonumber \\ 
m\ddot{y}_{i}&=&-\eta \dot{y}_i+\frac{q^{2}}{\epsilon}\sum_{i,j} F_{y,ij}+F_{y,i}^{T},
\label{MotionEq}
\end{eqnarray}
where $m$ and $q$ stand for the mass and charge of an electron (or, in general, a charged particle), 
$\eta$ is an effective viscous damping constant, 
and $\epsilon$ is the dielectric constant of the medium. 
The electron-electron interaction is modeled by the screened Coulomb potential 
(see, e.g.,~\cite{Bedanov,Piacente2004,Piacente2005}). 
Then the interparticle interaction force can be written as: 
\begin{equation}
\vec{F}_{ij}^{\alpha}=
-\frac{\partial}{\partial \alpha}
\frac{exp(|\vec{r}_{i}-\vec{r}_{j}|/\lambda)} 
{|\vec{r}_{i}-\vec{r}_{j}|} 
\label{ParticlesInter}, 
\end{equation}
where $\vec{r}_{i}=\left(x_{i},y_{i} \right)$, 
$\lambda$ is the screening length, 
and $\alpha$ is $x$ or $y$. 
Here 
$F_{dr}$ is the driving force 
(applied along the channel, i.e., in the $x$-direction), 
and 
$\vec{F}_{i}^{T}$ is a random thermal force 
obeying the following conditions: 
%\begin{equation}
$$
\langle F_{i}^{T}(t) \rangle = 0
$$
and
$$
\langle F_{i}^{T}(t)F_{j}^{T}(t^{\prime}) \rangle = 2 \,  \eta \,  k_{B} \,  T \,  \delta_{ij} \,  \delta(t-t^{\prime}). 
$$
It is convenient to choose $\lambda$ as a unit length. 
Then 
Eqs~(\ref{MotionEq}) can be rewritten in 
standard (see, e.g.,~\cite{Bedanov}) 
dimensionless form, 
\begin{eqnarray}
\ddot{x}_{i}&=&-\dot{x}_i+\sum_{i,j} f_{x,ij}+f_{x,i}^{T}+f_{dr} \nonumber \\
\ddot{y}_{i}&=&-\dot{y}_i+\sum_{i,j} f_{y,ij}+f_{y,i}^{T}, 
\label{MotionEqDL}
\end{eqnarray}
where we used the transformation 
$x^{\prime}=x/\lambda$, 
$t^{\prime}=t/t_{0}$ 
(and omit primes in the dimensionless equations) 
and the units of time and force, 
$t_{0}=m^{1/2}\lambda$ and $f_{0}=q^{2}/\lambda^{2}\epsilon$. 
In addition, without loss of generality, the dimensionless viscosity is set to unity in our simulations. 

The interaction of particles with the walls of the channel is modeled as hard-wall. 
The geometry of the channel and the constriction are defined by the parameters $a$, $b$, which are a half-length and half-width of the channel, respectively, and $l$, $w$ which are a half-length and half-width of the constriction, respectively, as shown in Fig.~\ref{ChannelSketch}. 
We impose periodic boundary conditions in the $x$-direction. 

\begin{figure*}[t]
\centerline{\epsfig{figure=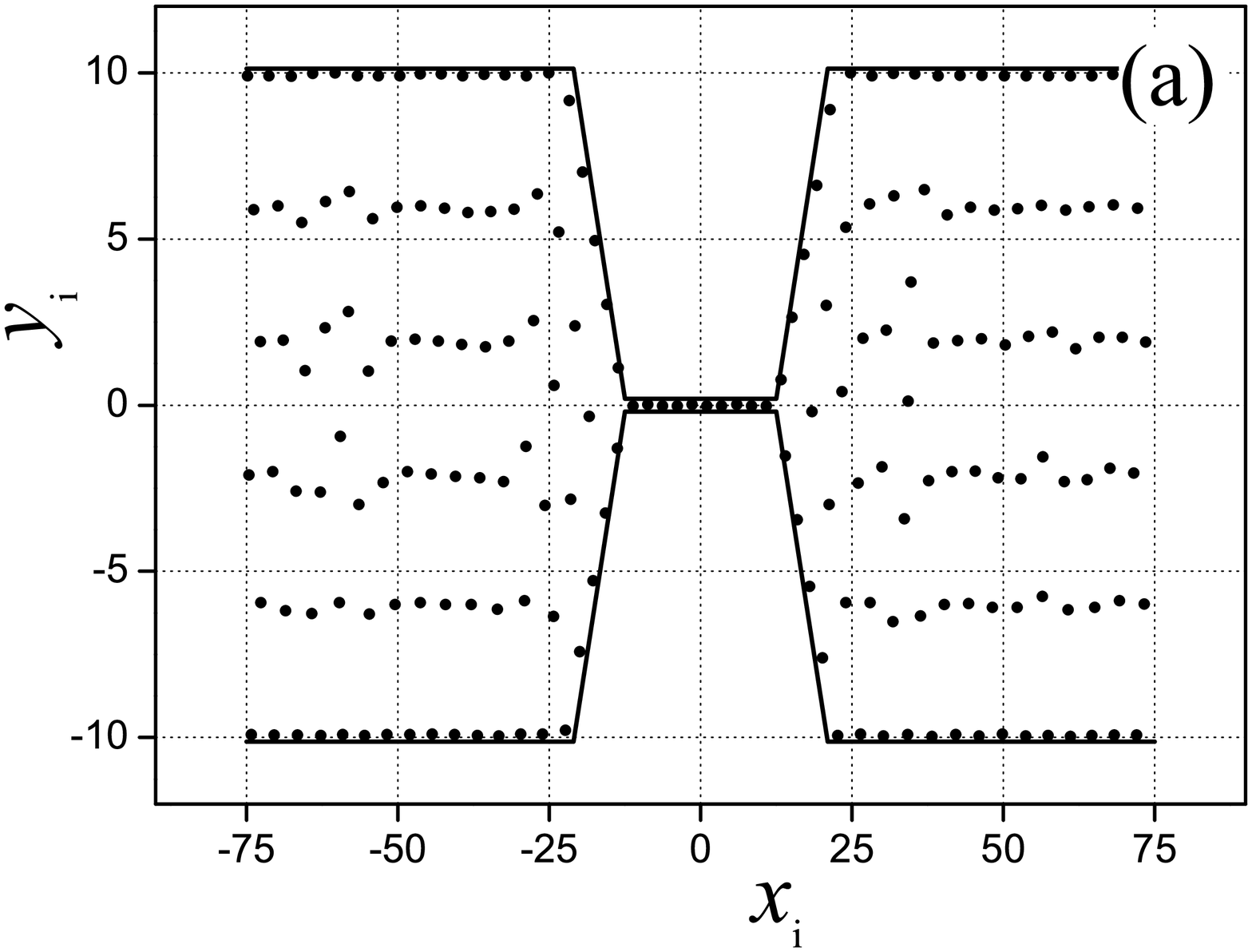,width=10cm}\hspace*{-1.6cm}\epsfig{figure=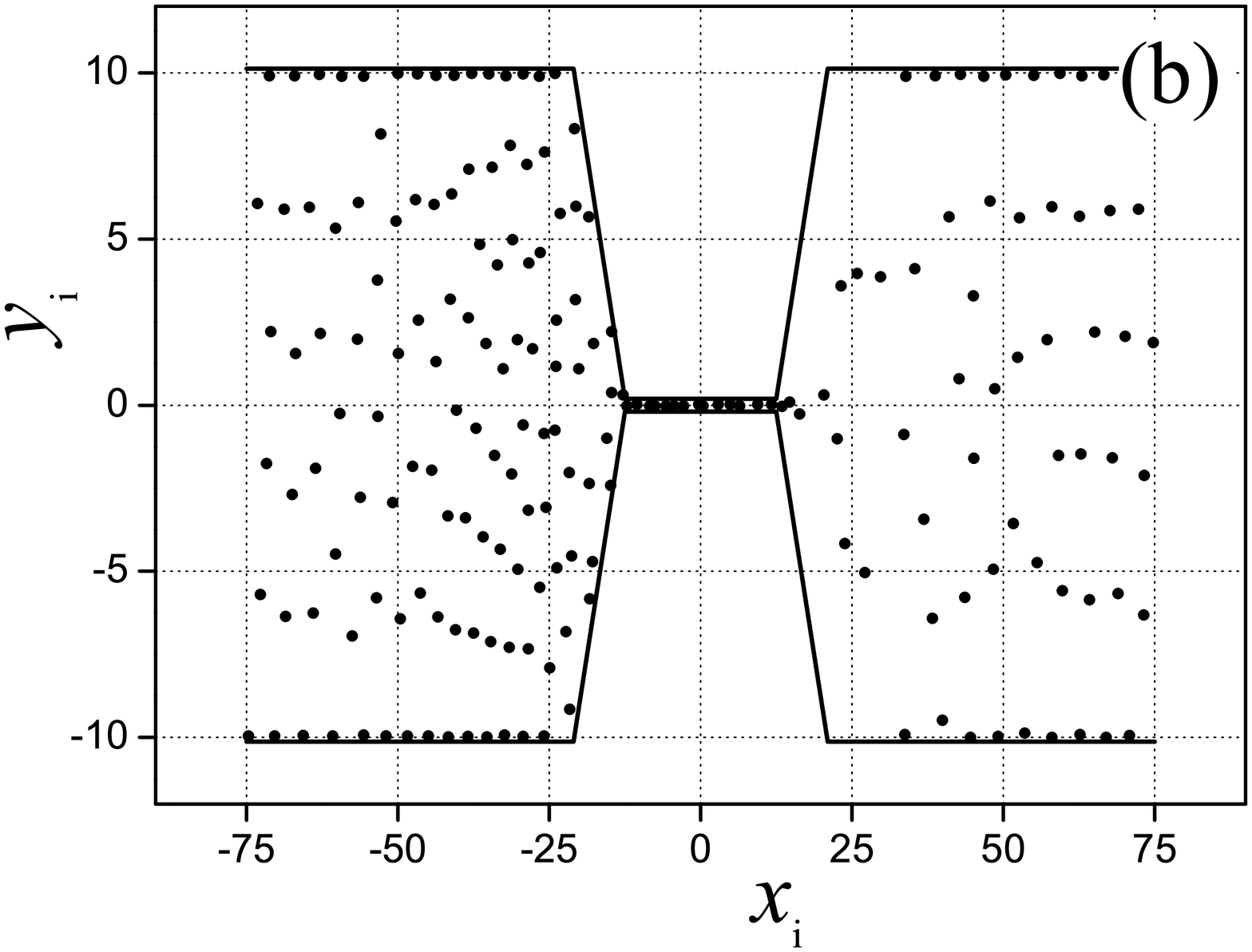,width=10cm}} \vspace*{-0.2cm}
\caption{Electrons distributions, 200 particles, $w=0.025$, (a) no driving force, (b) $f_{x}=0.2$.}
\label{1rowExL5}
\end{figure*}

\begin{figure*}
\centerline{\epsfig{figure=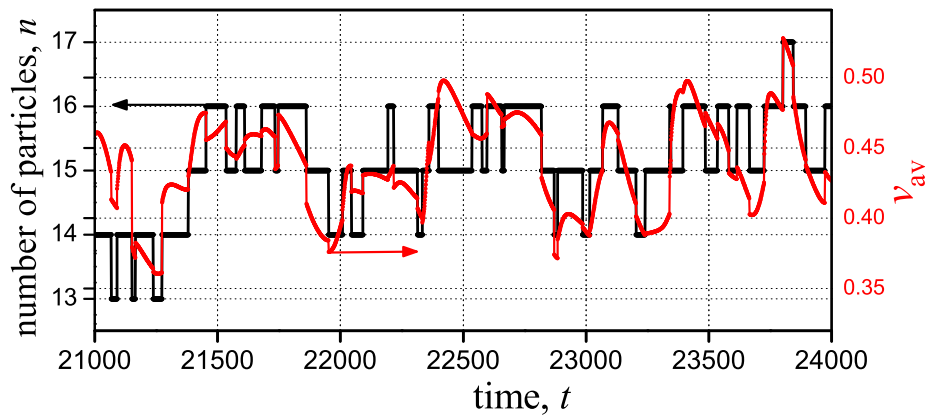,width=14cm}}
\vspace*{-5mm}
\caption{
Graphs of the average velocity of the electrons in the $x$-direction in the constriction (red curve) and the number of the electrons in the constriction (black curve), $w=0.025$, $f_{x}=0.2$.
}
\label{NumpVeloc}
\end{figure*}

Using computational methods of molecular dynamics, we investigate the transport of electrons for different constrictions and model parameters. 
The initial state of the system is prepared using simulated annealing simulation (SAS). 
For this purpose, we initially set some value of temperature which is high enough to avoid trapping of the system in a metastable state, and then gradually decrease the temperature. The obtained initial state then is either the ground state of the system or a low-energy state close to the ground state. 
To study the dynamics of the particles, we apply a weak driving force which can be obtained for each specific constriction and particle configuration from the corresponding $IV$-curve (i.e., the average velocity versus driving force curve), similar to the experimental procedure.

\section{Electron transport in a channel with a symmetric constriction}
  
In this section, we consider channels with different types of constrictions. 
Let us start with investigating the dynamics of electrons in a  channel with a constriction with inclined boundaries forming an angle of 45$^\circ$ with the lateral boundaries of the channel (see Fig.~\ref{ChannelSketch}). 
The electrons form the Wigner crystal structure inside the channel, and when we decrease the constriction width, the number of the electrons rows in the constriction decreases. 
Ultimately for very narrow constriction the regimes of ``zigzag'' and ``quantum wire'' when only two or one row of the particles is formed in the constriction (i.e., the {\it single-file} regime) can be observed. 
Consequently, density of the electrons and the number of rows in the rest of the channel grow as we decrease the width of the constriction. 

Our goal is to study the quasi-one-dimensional transport with additional driving force applied in the $x$-direction $f_{x}$. 
Therefore, we apply the driving force in the $x$-direction and consider very narrow constriction in the channel such that the electrons pass the constriction in one row. 
Examples of the electron distributions in the channel with constriction are shown in Fig.~\ref{1rowExL5}(a) (when no driving force is applied) 
and in Fig.~\ref{1rowExL5}(b) (when a weak driving is applied leading to a transport current). 
As one can see in the first figure (Fig.~\ref{1rowExL5}(a)), the particles are ordered in rows both to the left and to the right from the constriction. 
The row structure is only destroyed near the constriction where the width of the channel is decreased. 
Turning on the driving force destroys the symmetric electron  distribution and creates gradients in the electron density (Fig.~\ref{1rowExL5}(b)). 
In order to remain close to the equilibrium state, the driving force therefore is chosen as weak as possible. 
This allows us to remain within the quasi-stationary regime and to model the experimental situation with a minimal transport current. 
This will allow us to achieve a better understanding of the phenomena observed in the 
experiments~\cite{Rees-Kono-2010,Araki-Hayakawa-2012,Rees-Totsuji-2012}.

\begin{figure}
\centerline{\epsfig{figure=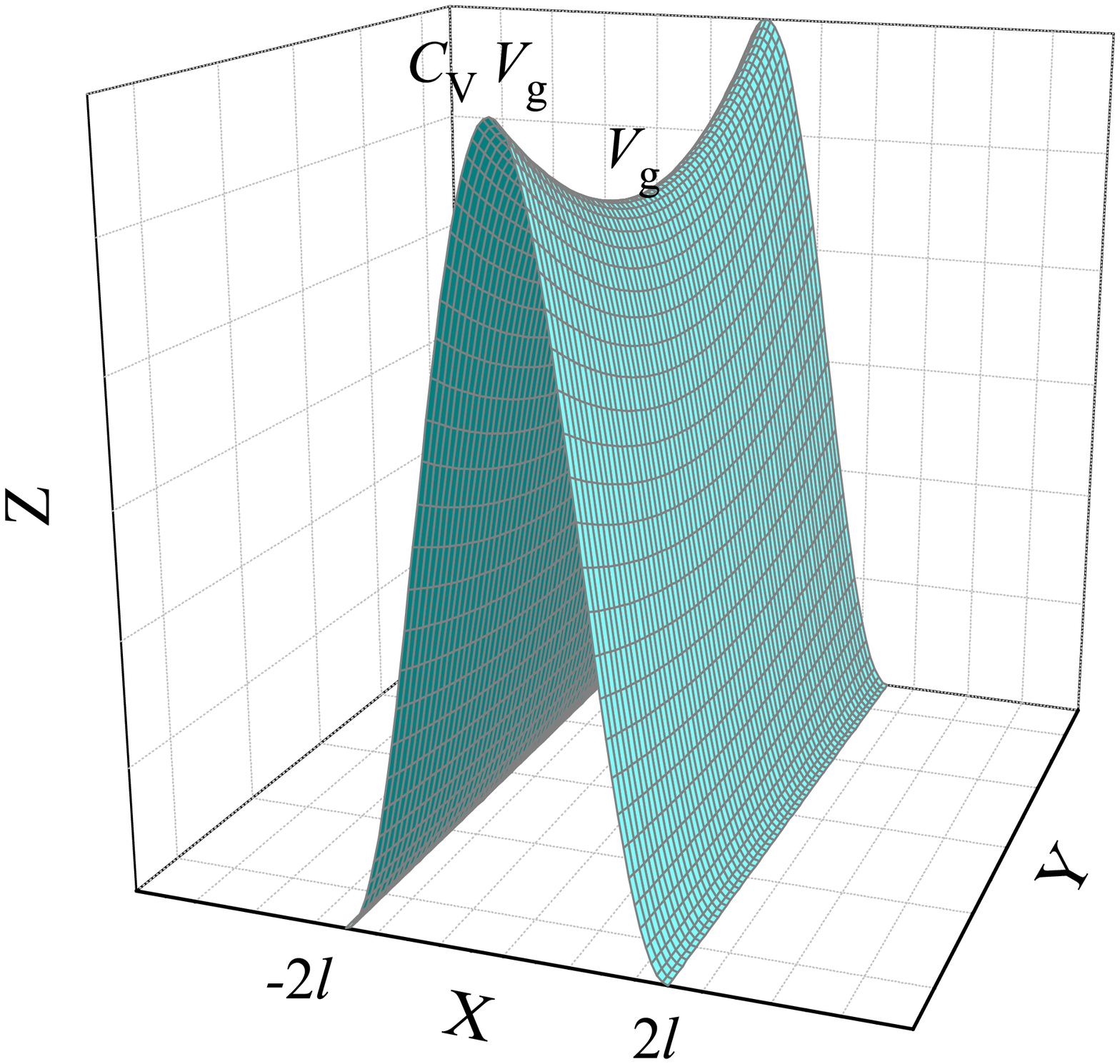,width=10cm}}
\vspace*{-0.5cm}
\caption{Sketch of the saddle-point potential in the channel.}
\label{SaddlePointPotential}
\centerline{\hspace*{5mm}\epsfig{figure=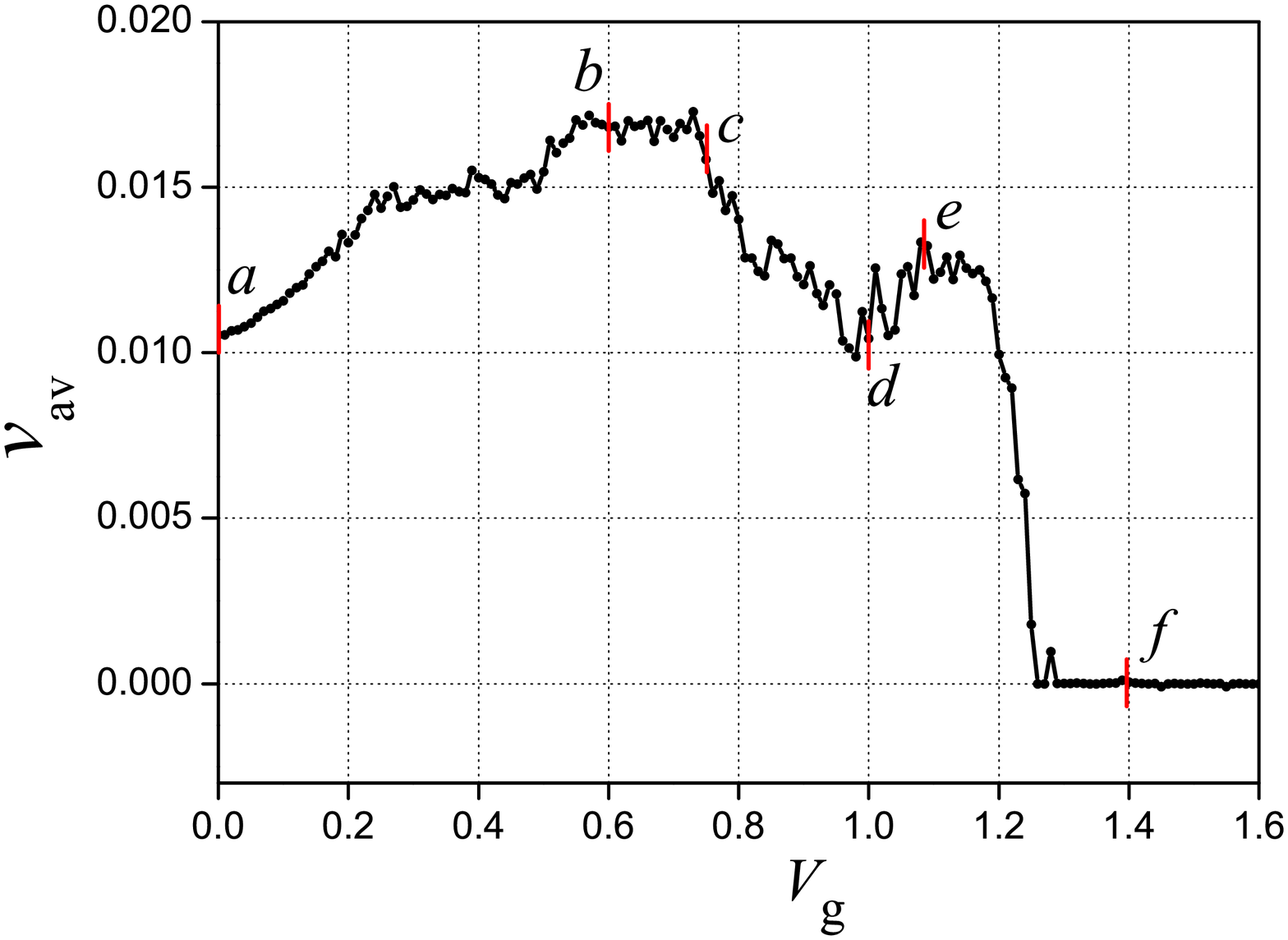,width=9.5cm}}
\vspace*{-0.2cm}
\caption{The average velocity of the particles in the $x$-direction in the constriction $V_{av}$ versus the gate voltage $V_{g}$, $f_{x}=0.015$.}
\label{MVfromVgSmooth}
\end{figure}

For a fixed value of the driving force $f_{x}$, the average 
velocity of the electrons in the constriction in the $x$-direction exhibits a correlation with the number of the electrons in the constriction. 
The oscillations in the average velocity are due to single-electron entry/exit in/from the constriction. 
The corresponding functions of the average velocity in the $x$-direction and the number of the particles versus iteration time are shown in Fig.~\ref{NumpVeloc}. 
As one can see from the plot, 
there is a pronounced correlation between the change in the number of particles in the channel and in the average velocity. 
Thus an increase (decrease) in the number of particles by one or two, is associated with an increasing (decreasing) jump in the average velocity. 
It is clear, however, that the life-time of a state with $N+1$ particles should be not too short to have an impact on the average velocity. 
Below, we assume some weak driving such that it provides electron transport through the constriction but does not significantly influence the equilibrium particle distributions. 

Let us now analyze how the electron transport through the channel with constriction can be influenced by changing driving force $f_{x}$, i.e., calculate the ``average velocity $\langle v \rangle$ vs. driving force $f_{dr}$'' which is the analog of the ``$IV$-curve'' of the system.

\subsection{Oscillations in $v_{av} - V_{g}$ curve} 

In this section, we address the phenomenon recently observed in the experiment~\cite{Rees-Totsuji-2012} where 
the conductance of the classical point contact revealed pronounced oscillations for small conductance values. 
The oscillations were observed for short constrictions while for long constrictions, instead, a monotonic behavior was revealed. 

The experiment~\cite{Rees-Totsuji-2012} suggests that in case of narrow constrictions the gate potential has a shape of a saddle point. 
In this section we consider a narrow channel with a constriction which is modeled as a saddle-point potential, as shown in Fig.~\ref{SaddlePointPotential}. 

In our model, the saddle-point potential is defined as follows: 
\begin{equation}
Z\left(x,y\right) = \frac{V_{g}}{2}\left( \frac{C_{V}-1}{b^{2}}y^{2}+1 \right)  
\left( \cos\left(\frac{\pi}{2l}x\right)+1 \right),
\label{SPPotential}
\end{equation}
where parameters $V_{g}$ and $C_{V}$ define the height of the potential:
$$\begin{array}{lcl}
Z\left(0,0\right) &=& V_{g},  \nonumber \\
Z\left(0,-b\right) &=& Z\left(0,b\right) = C_{V}V_{g}.  \nonumber
\end{array}$$
We choose the potential to extend over the interval $\left( -2l; 2l \right)$ and we calculate the average velocity of the particles in the interval $\left( -l; l \right) $. 
To be consistent with the experimental situation~\cite{Rees-Totsuji-2012}, we choose very low values of the driving force $f_{x}$ such that they do not destroy the equilibrium Wigner crystal configurations but provide the motion of electrons through the constriction.

\begin{figure*}[t!]
\centerline{\epsfig{figure=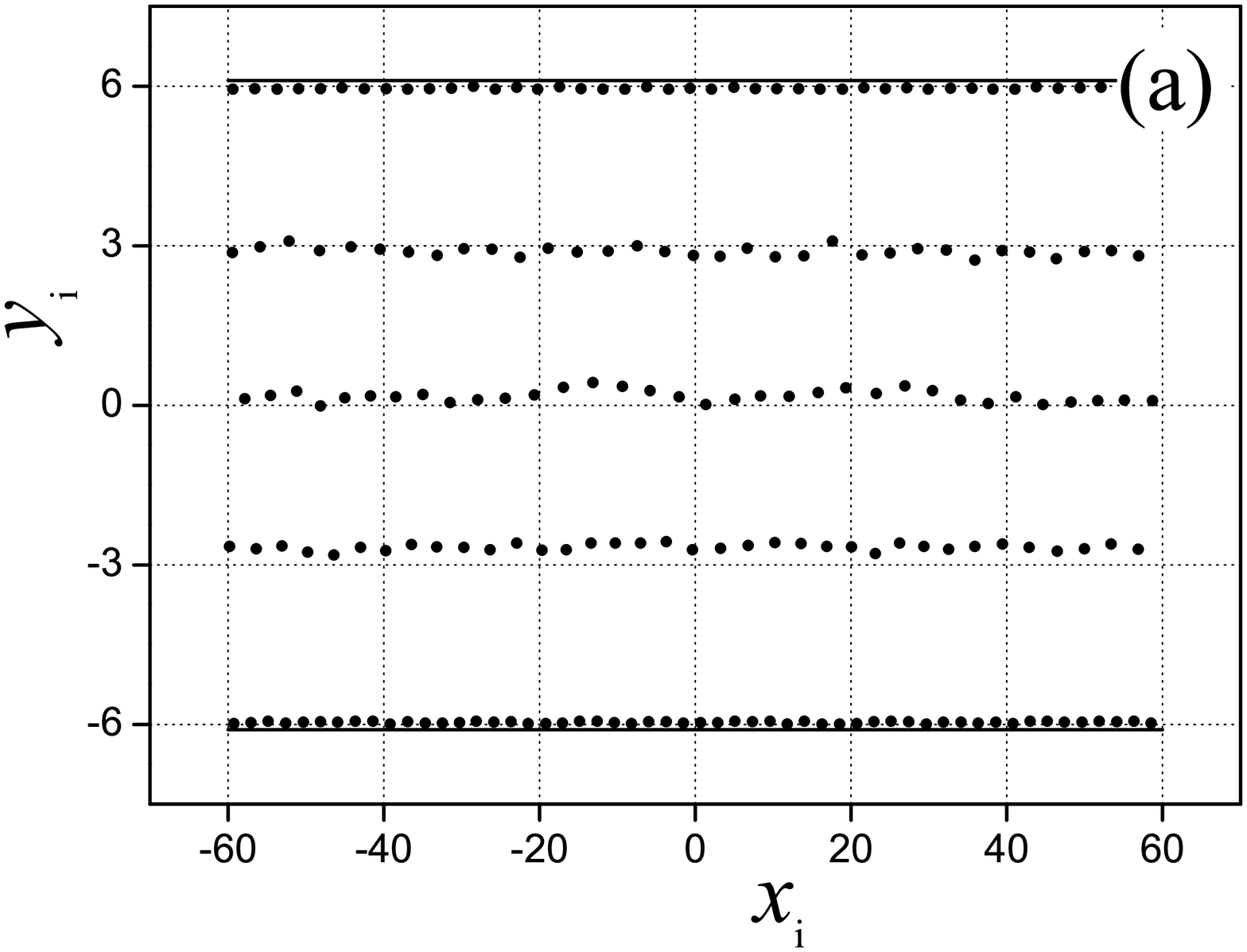,width=9cm}\hspace*{-1.5cm}\epsfig{figure=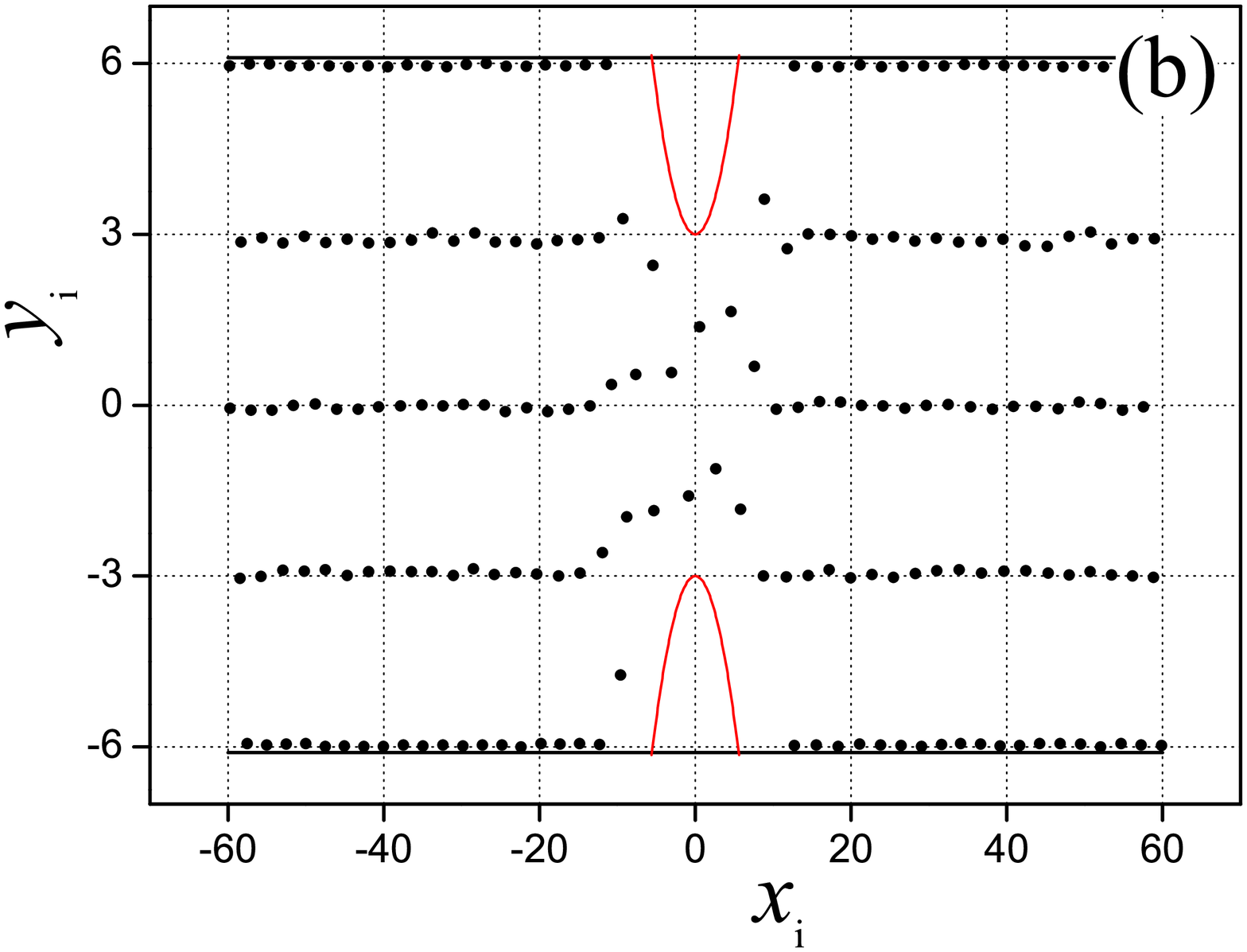,width=9cm}}
\vspace*{-0.6cm}
\centerline{\epsfig{figure=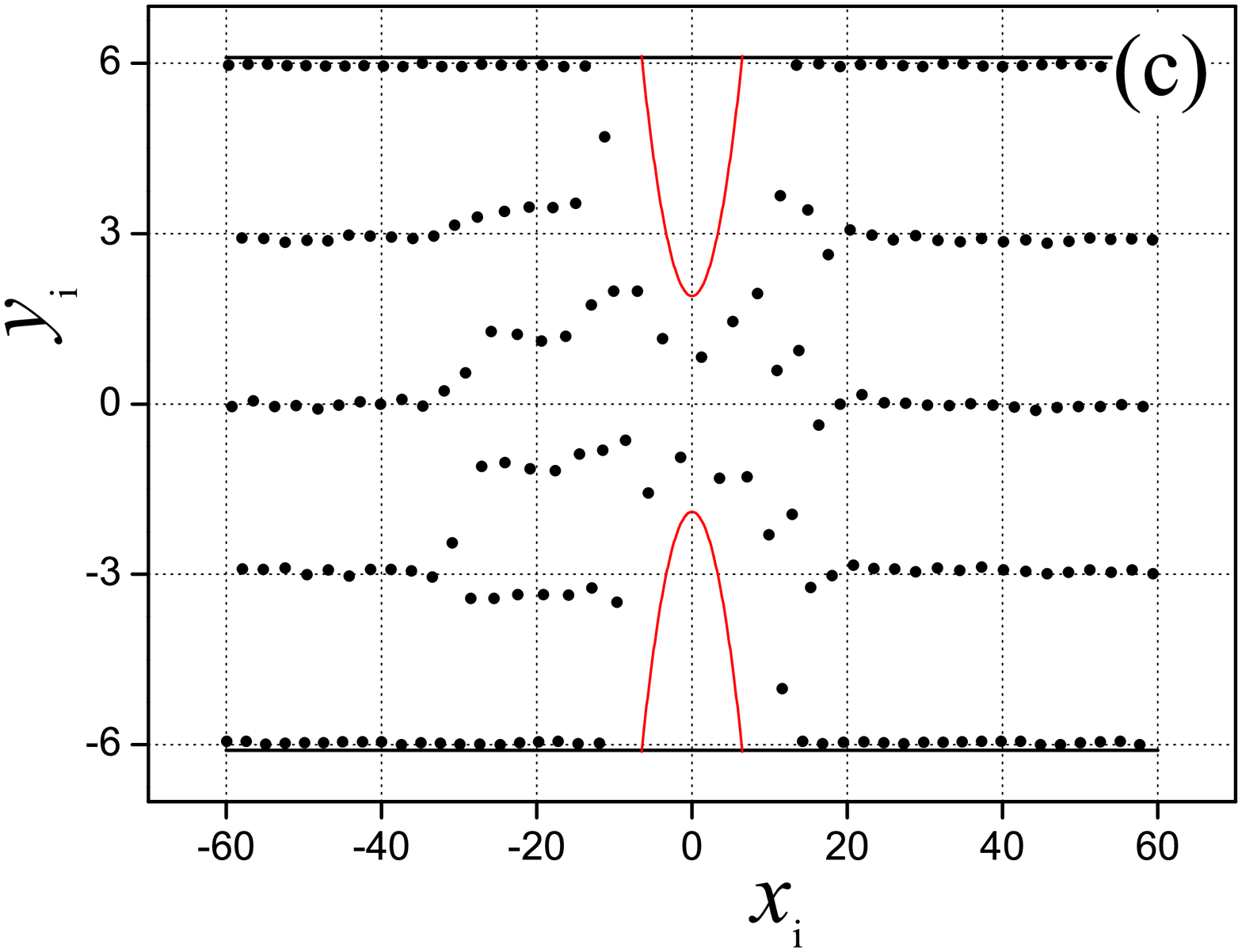,width=9cm}\hspace*{-1.5cm}\epsfig{figure=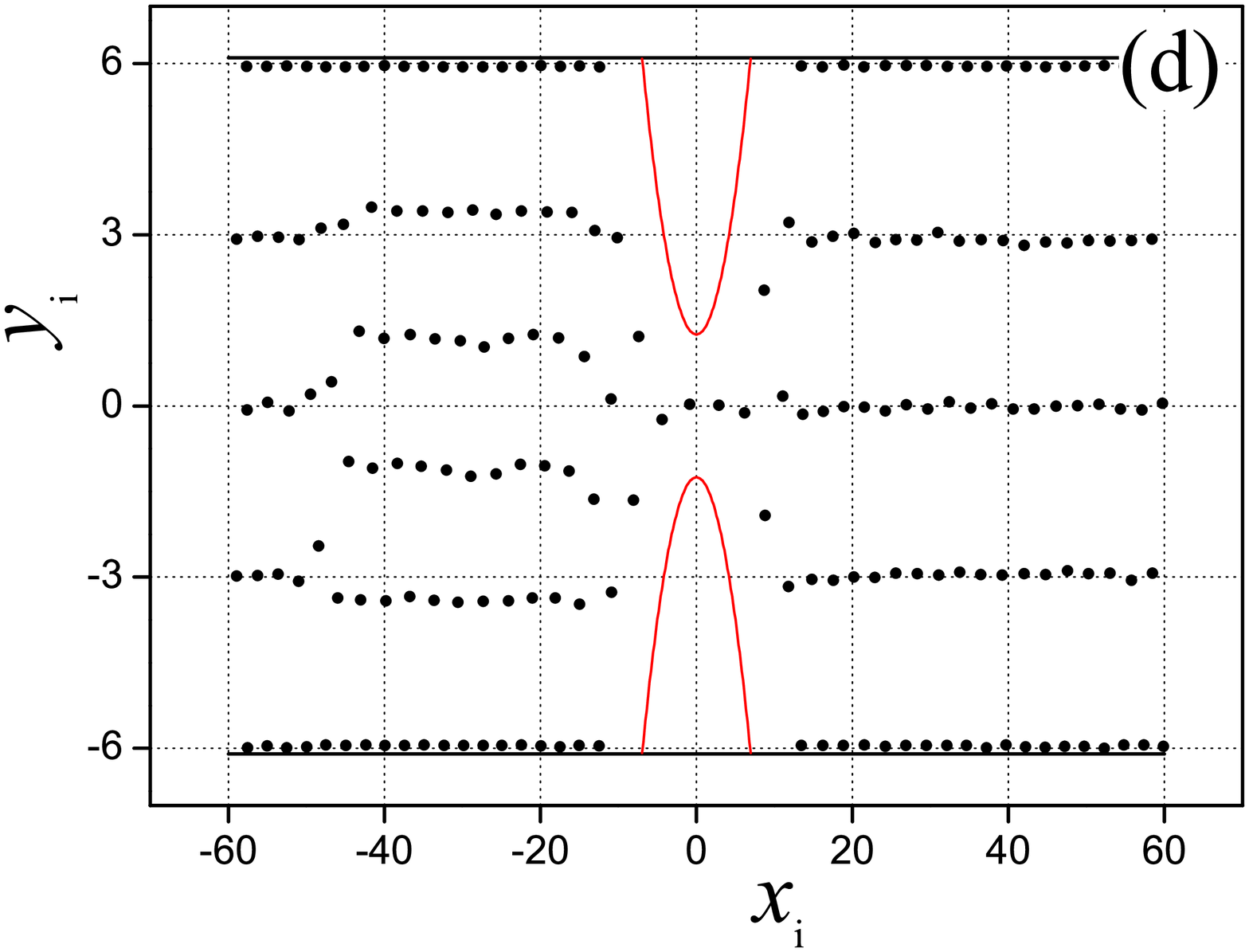,width=9cm}} 
\vspace*{-0.6cm}
\centerline{\epsfig{figure=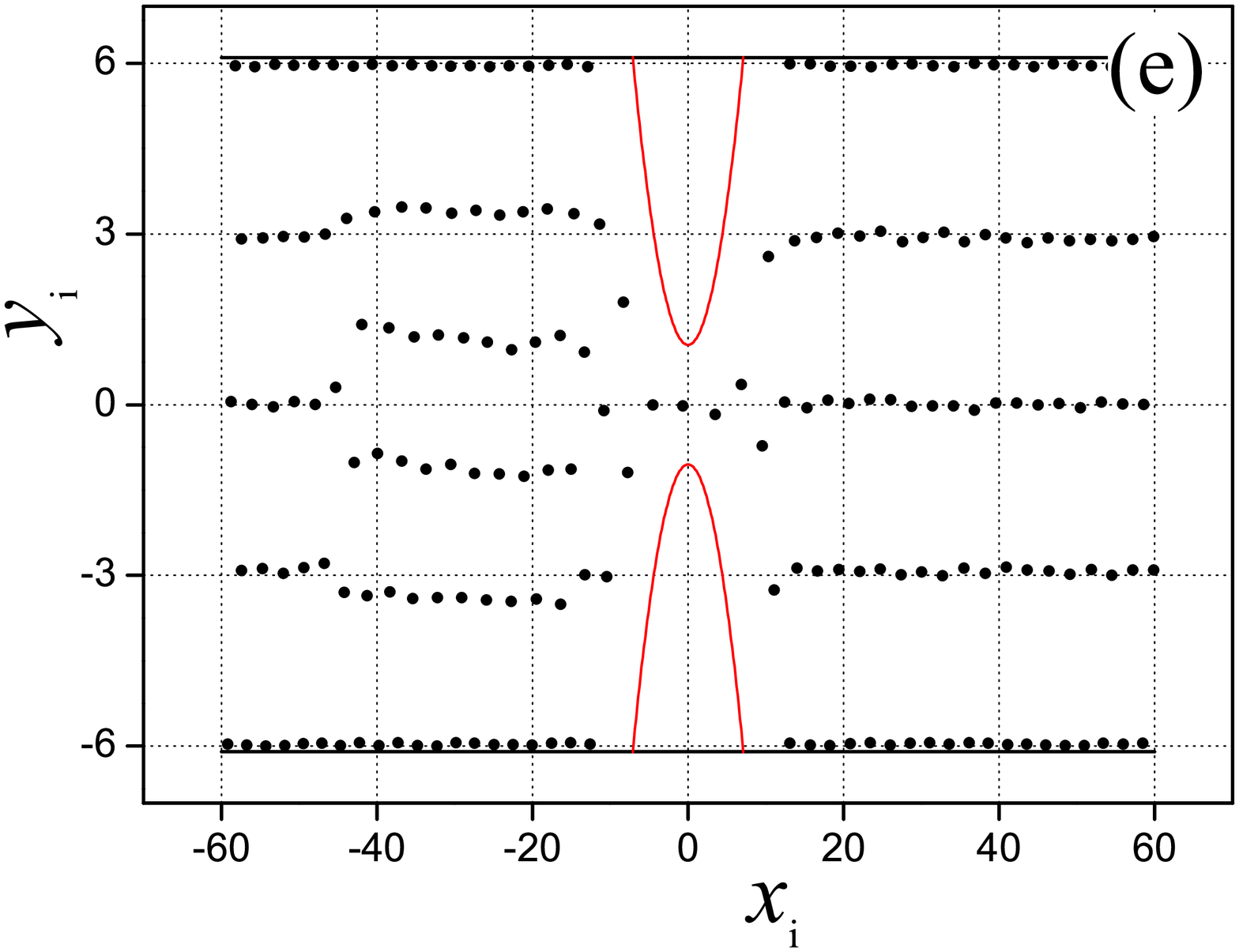,width=9cm}\hspace*{-1.5cm}\epsfig{figure=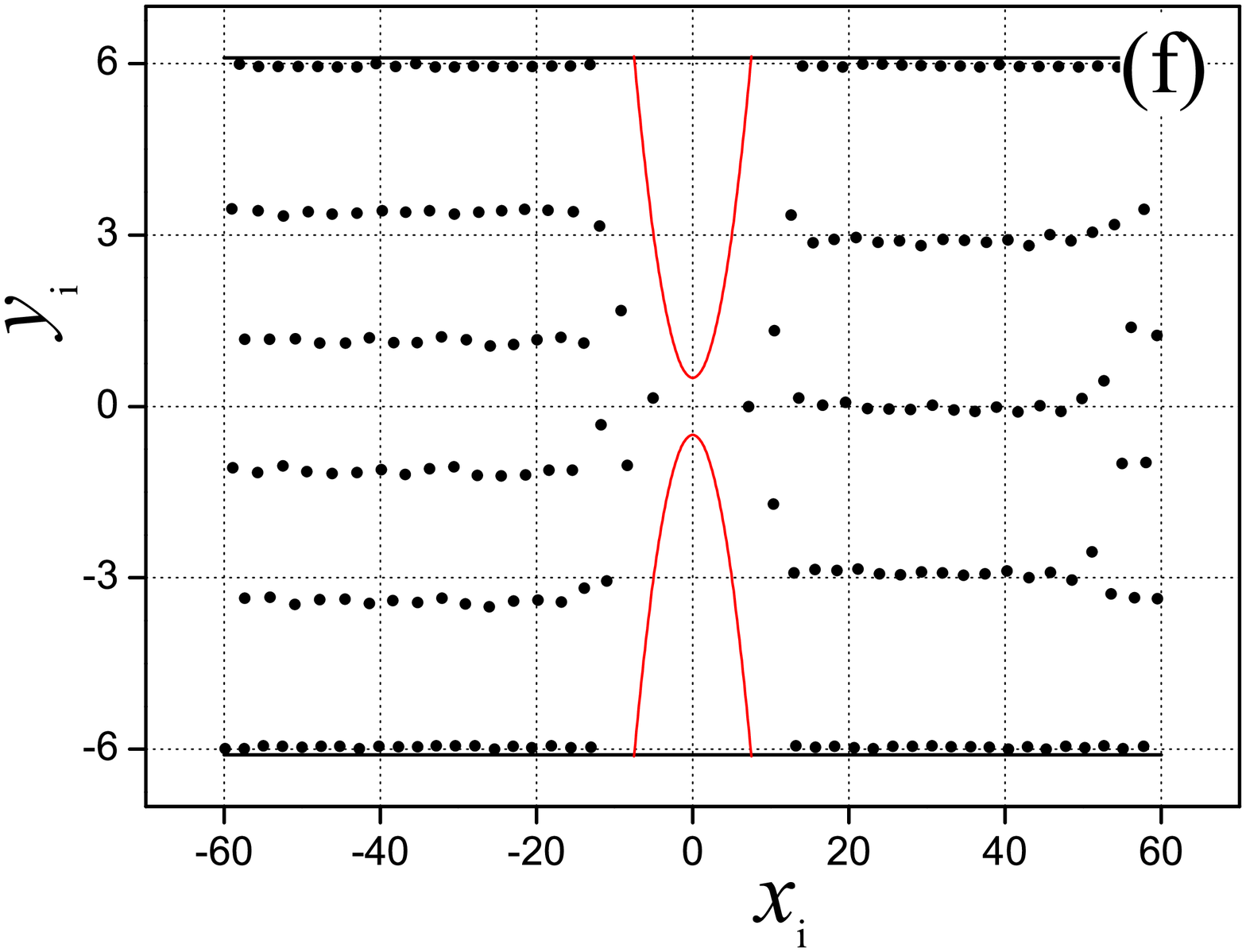,width=9cm}}
\vspace*{-0.3cm}
\caption{
Electron distributions in the channel with a saddle-point potential, for different values of the potential parameter $V_{g}$: (a) 0.0, (b) 0.6, (c) 0.75, (d) 1.0, (e) 1.08,  (f) 1.4.
}
\label{SnShL2F015}
\end{figure*}
  
First, we consider a relatively short constriction with length $l=8$. 
Fig.~\ref{MVfromVgSmooth} shows the resulting average velocity as a function of the gate voltage~\cite{voltage}, 
while the electron distributions corresponding to the values of $V_{g}$ marked with characters (a to f), are shown in Fig.~\ref{SnShL2F015}. 

\begin{figure}[t!]
\centerline{\epsfig{figure=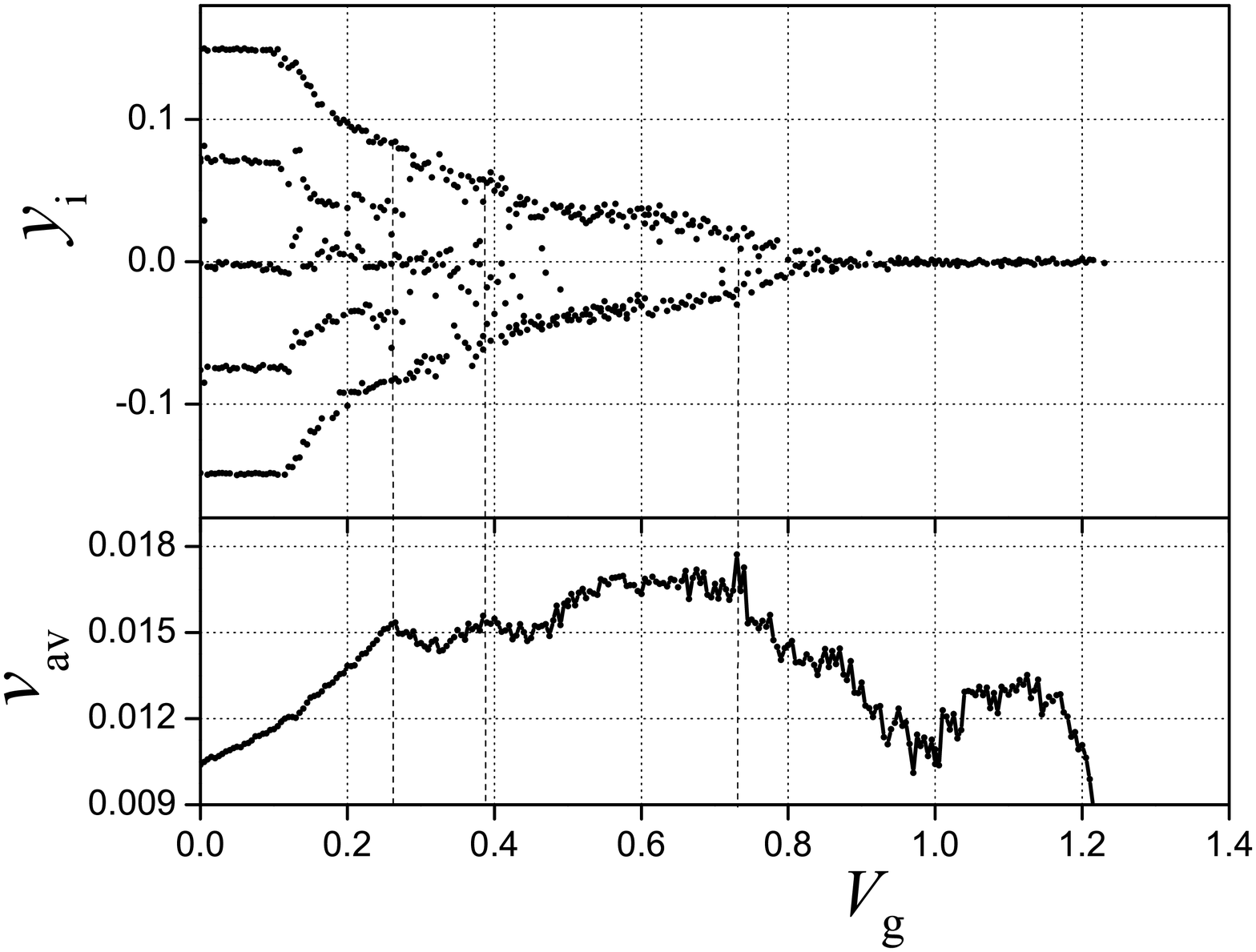,width=9cm}}
\caption{The average velocity of the particles in the $x$-direction in the constriction as defined by 
Eq.~(\ref{SPPotential}) (bottom panel), and ordinates of the electrons $y_{i}$ for the moments when they pass the center of the channel (top panel); $N=200$ particles, $f_{x}=0.015$.}
\label{BifDL2F015}
\end{figure}  

\begin{figure}[]
\centerline{\epsfig{figure=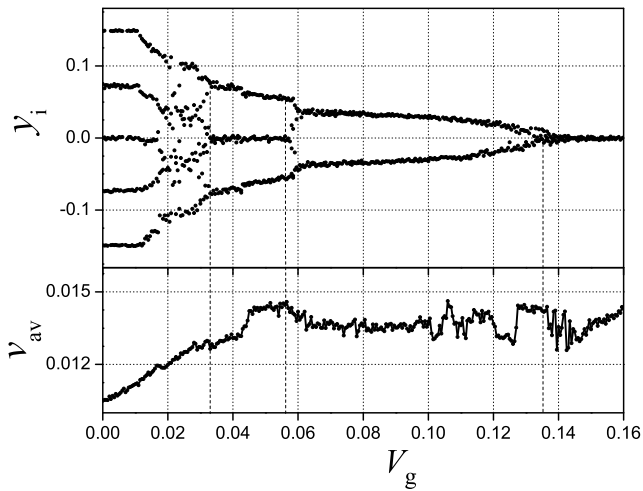,width=9cm}}
\caption{Graph of the average $x$-direction velocity $v_{av}$ of the particles in the long constriction and ordinates of the electrons $y_{i}$ in the moments when they pass the center of the channel; $N=200$ particles, $f_{x}=0.015$.}
\label{BifDL5F015}
\end{figure}

\begin{figure}[]
\centerline{\epsfig{figure=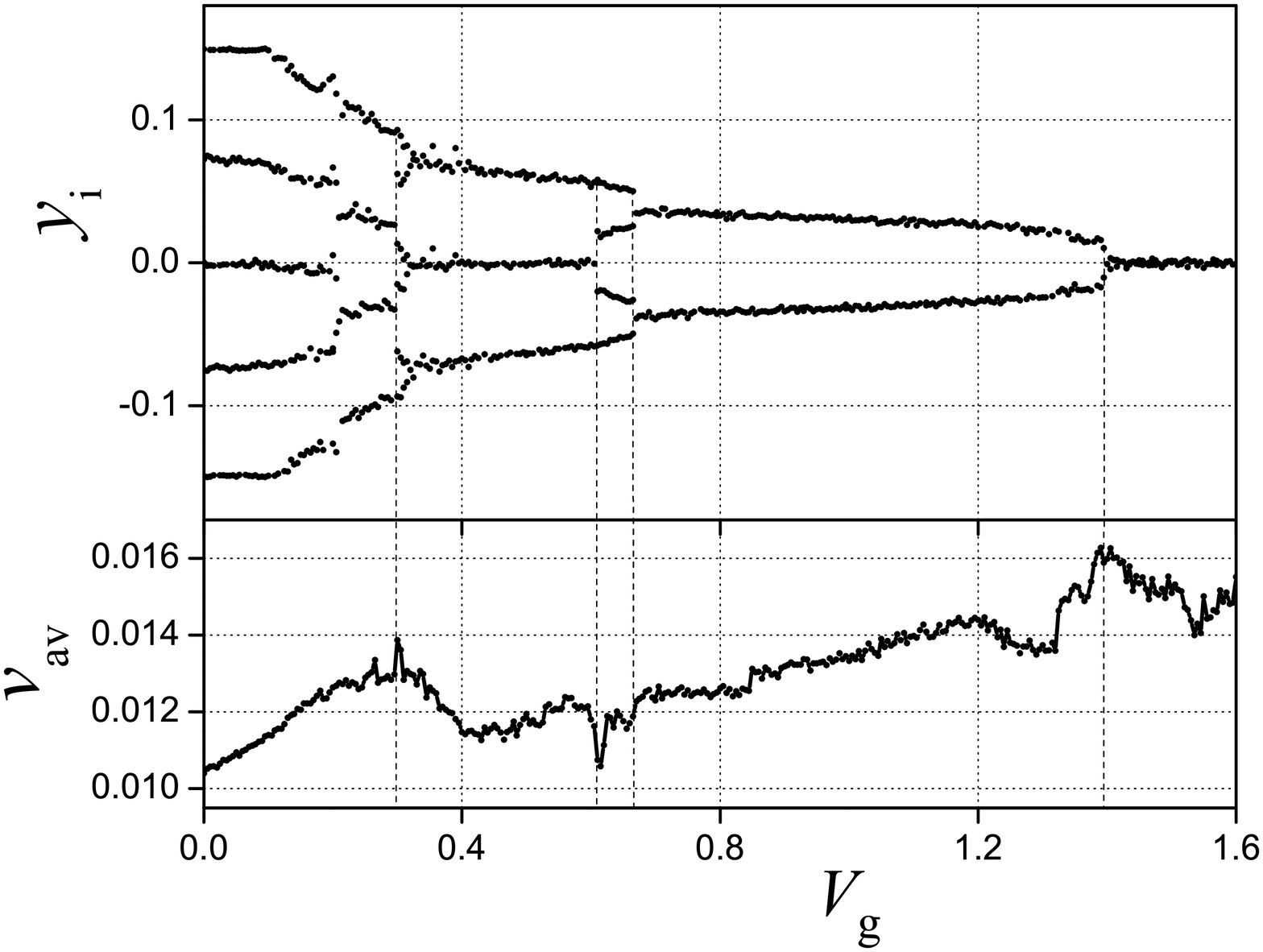,width=9cm}}
\caption{The average velocity of the particles in the $x$-direction in the modified constriction as defined by Eq.(\ref{SPPlatoPot}) (bottom panel), and ordinates of the electrons $y_{i}$ in the moments when they pass the center of the channel (top panel); $N=200$ particles, $f_{x}=0.015$.}
\label{BifDL4Lp4F015}
\end{figure}

The analysis of the electron distributions indicates that the oscillations in the average velocity curve are caused by the changing number of rows of electrons in the constriction. 
Thus the first and the second peaks in the 
$v_{av}(V_{g})$-curve 
correspond to 
one (Fig.~\ref{SnShL2F015}(e)) 
and two (Fig.~\ref{SnShL2F015}(c))  
rows of electrons moving through the constriction. 
Clearly, in case of a saddle-point constriction where the electrons are distributed generally non-uniformly along the $x$-direction, these rows are best defined in {\it very short} constrictions, there they are formed just by a few electrons  (or ultimately just by one electron). 
Therefore, a change in their number results in a pronounced oscillation in the average velocity of electrons. 

In order to obtain a better insight in the transitions between the dynamical regimes with different numbers of the rows, we track the $y$-coordinates of the electrons passing through the center of the constriction at $x=0$ during some time interval, e.g., $t=5000 \Delta t$, for gradually changing $V_{g}$. 
The results are summarized in Fig.~\ref{BifDL2F015}.
From these plots one can see that the transition points between the states with different numbers of rows in the constriction correspond to the peaks of the average velocity curve (lower panel).

\subsection{Suppressed oscillations in long constriction} 

For long constrictions there are no well pronounced oscillations unlike in the above case of shorter constrictions. 
Although the transitions between the states with different numbers of rows are still observed, they do not result in pronounced features in the velocity curve.
The reason for this behavior is that the particle distributions in longer saddle-point constrictions are not formed by rows which are uniform along the constriction and thus cannot be characterized by a unique number of rows. 
Instead, the number of rows varies along the length of the constriction resulting in a smooth monotonic change of 
$v_{av}$ versus $V_{g}$, as shown in Fig.~\ref{BifDL5F015}. 
This result is in agreement with the observations of the 
experiment~\cite{Rees-Totsuji-2012}.

\subsection{Modified potential}

Next, we consider a slightly modified model potential such  that the central part of the constriction contains a short 
{\it plateau} thus stabilizing the rows: 
\begin{equation}
Z\left(x,y\right) = 
\begin{cases}    
\frac{V_{g}}{2}\left( \frac{C_{V}-1}{b^{2}}y^{2}+1 \right)  
\left( \cos\left(\frac{\pi}{2l}\left(x+l\right)\right)+1 \right), \ x<-l \\      
V_{g}\left( \frac{C_{V}-1}{b^{2}}y^{2}+1 \right), \ \ \ \ \ \ \ \ \ \ \ \ \ \ \ \ \ \ \ \ \ \ -l \leq x \leq l \\
\frac{V_{g}}{2}\left( \frac{C_{V}-1}{b^{2}}y^{2}+1 \right)  
\left( \cos\left(\frac{\pi}{2l}\left(x-l\right)\right)+1 \right), \ x>l     
\end{cases} \label{SPPlatoPot}
\end{equation}

\begin{figure*}
\centerline{\epsfig{figure=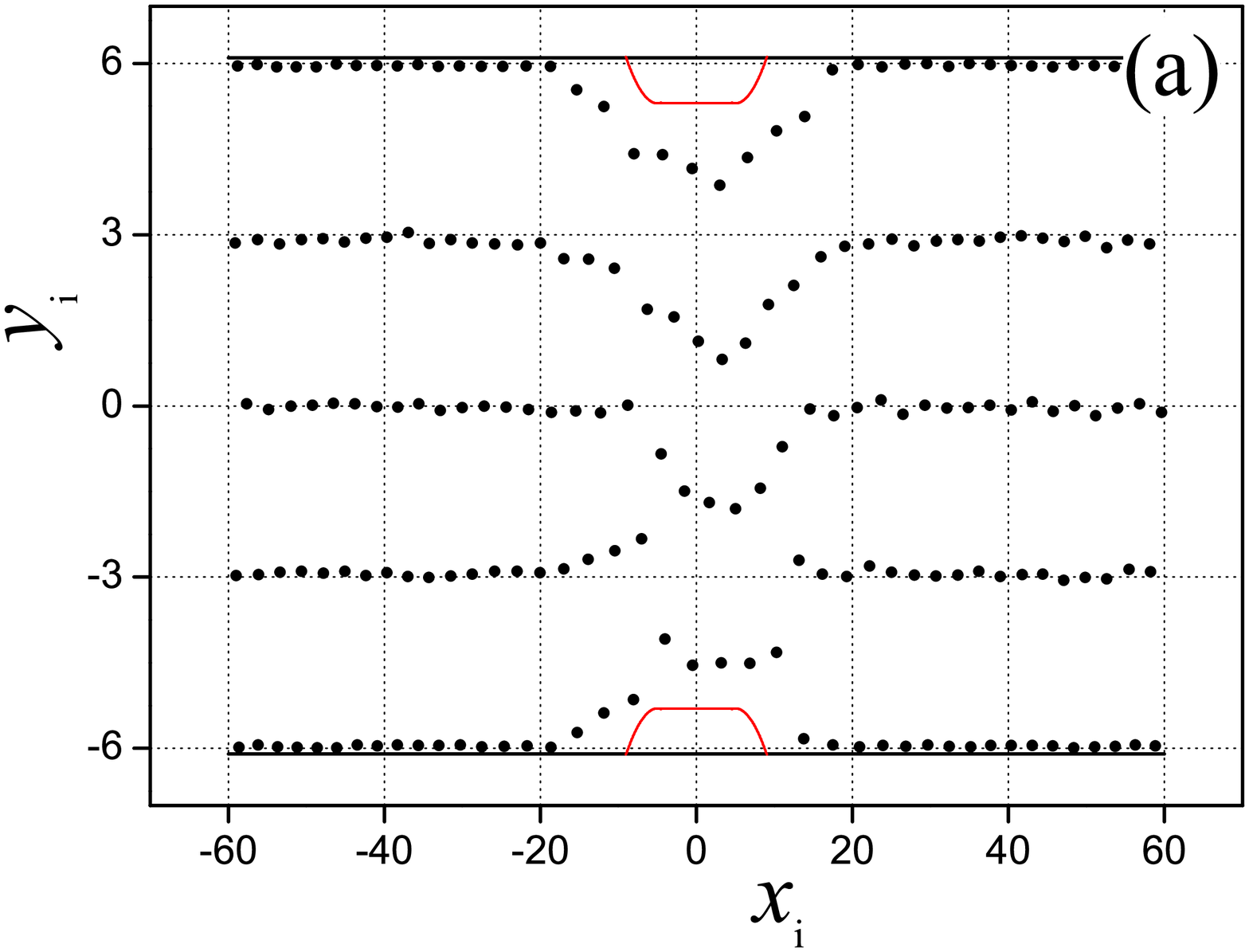,width=9cm}\hspace*{-1.2cm}\epsfig{figure=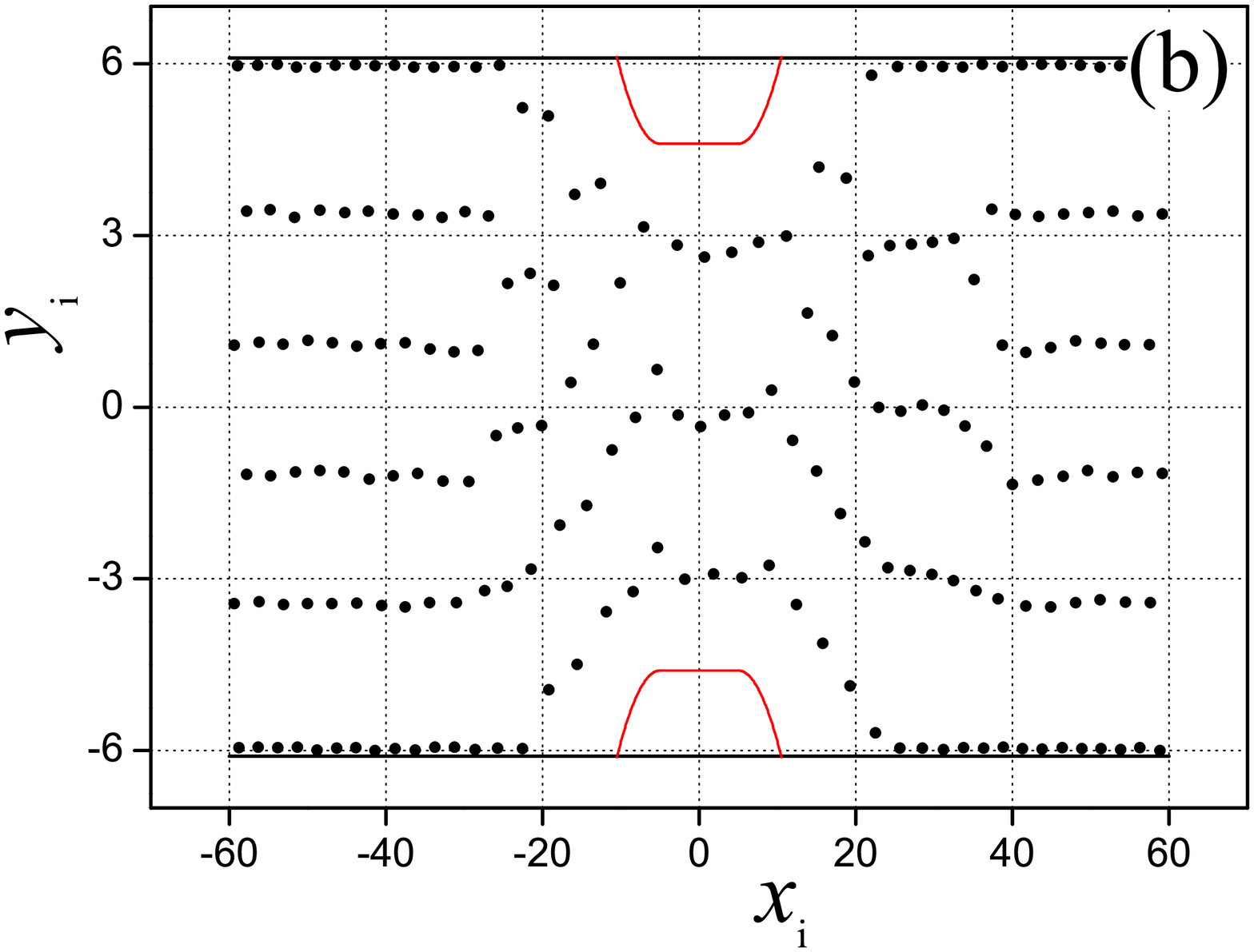,width=9cm}}
\vspace*{-0.4cm}
\centerline{\epsfig{figure=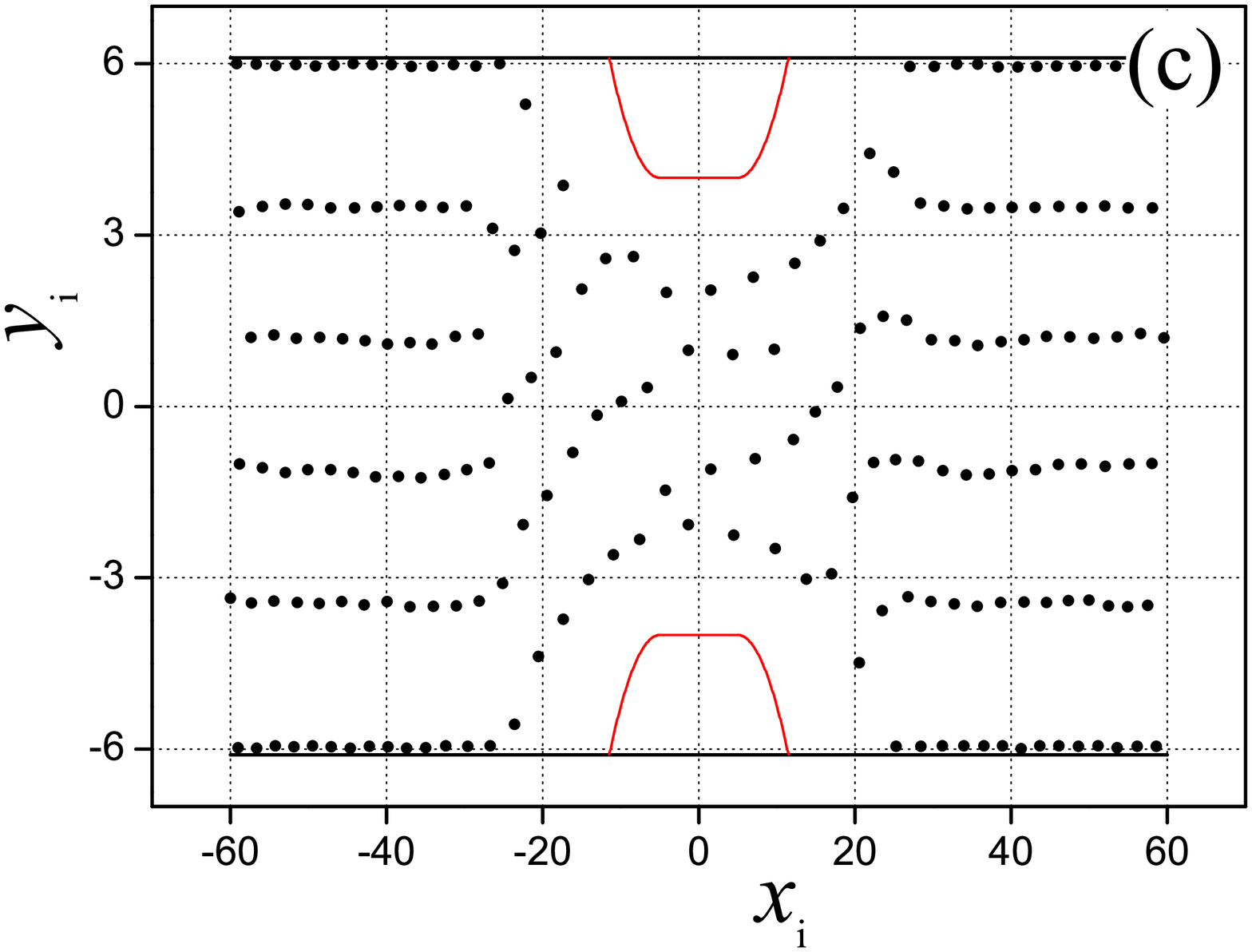,width=9cm}\hspace*{-1.2cm}\epsfig{figure=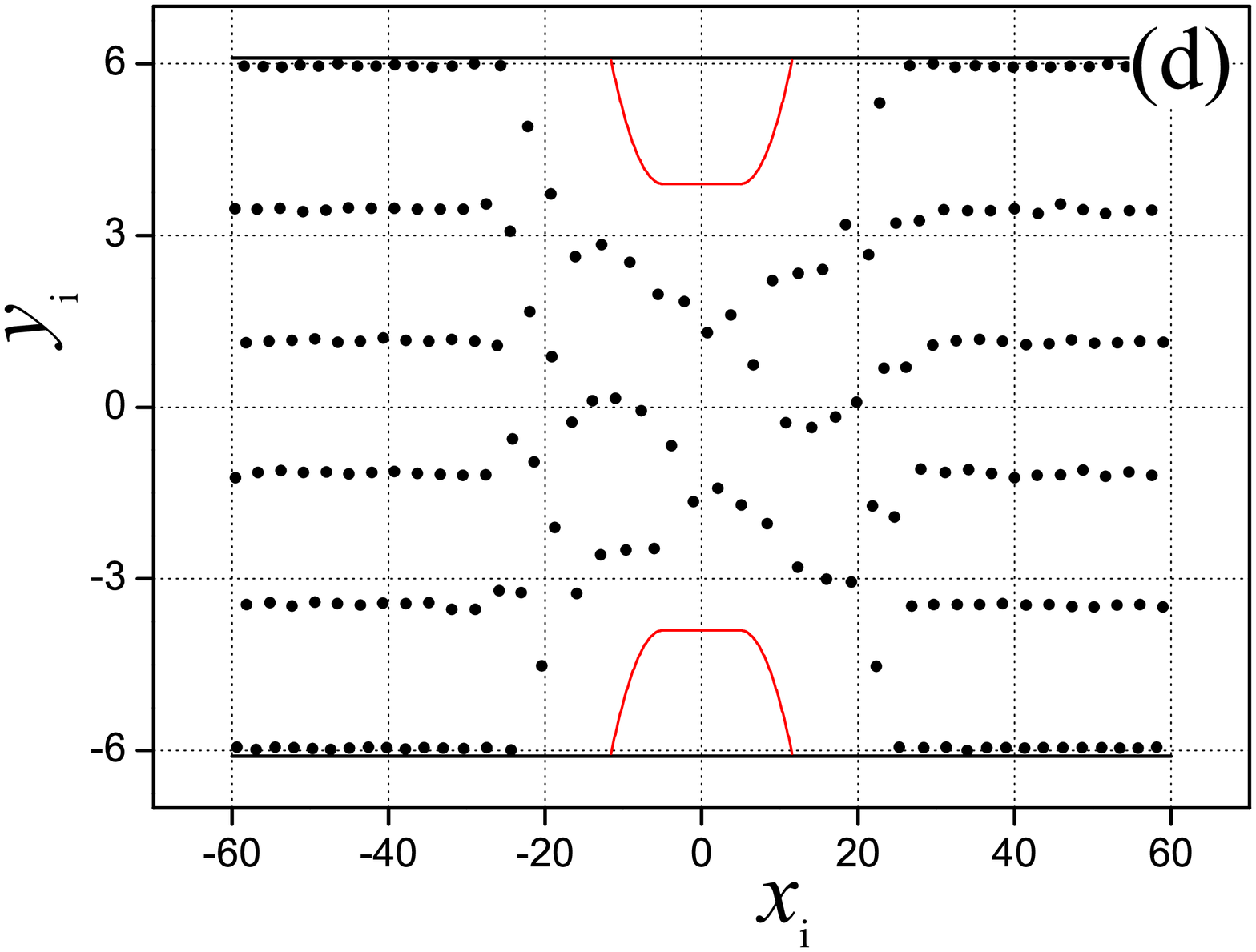,width=9cm}} 
\vspace*{-0.4cm}
\centerline{\epsfig{figure=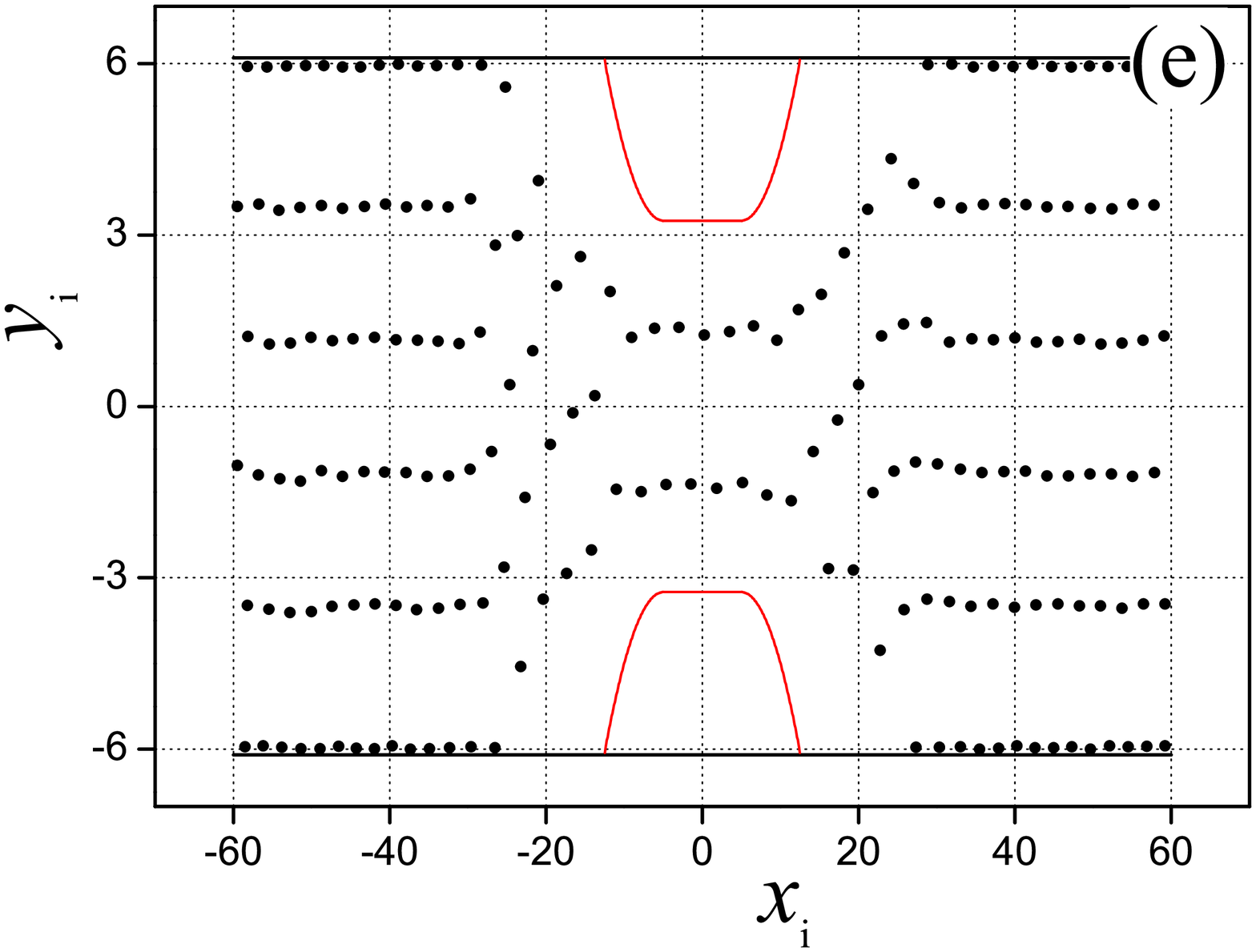,width=9cm}\hspace*{-1.2cm}\epsfig{figure=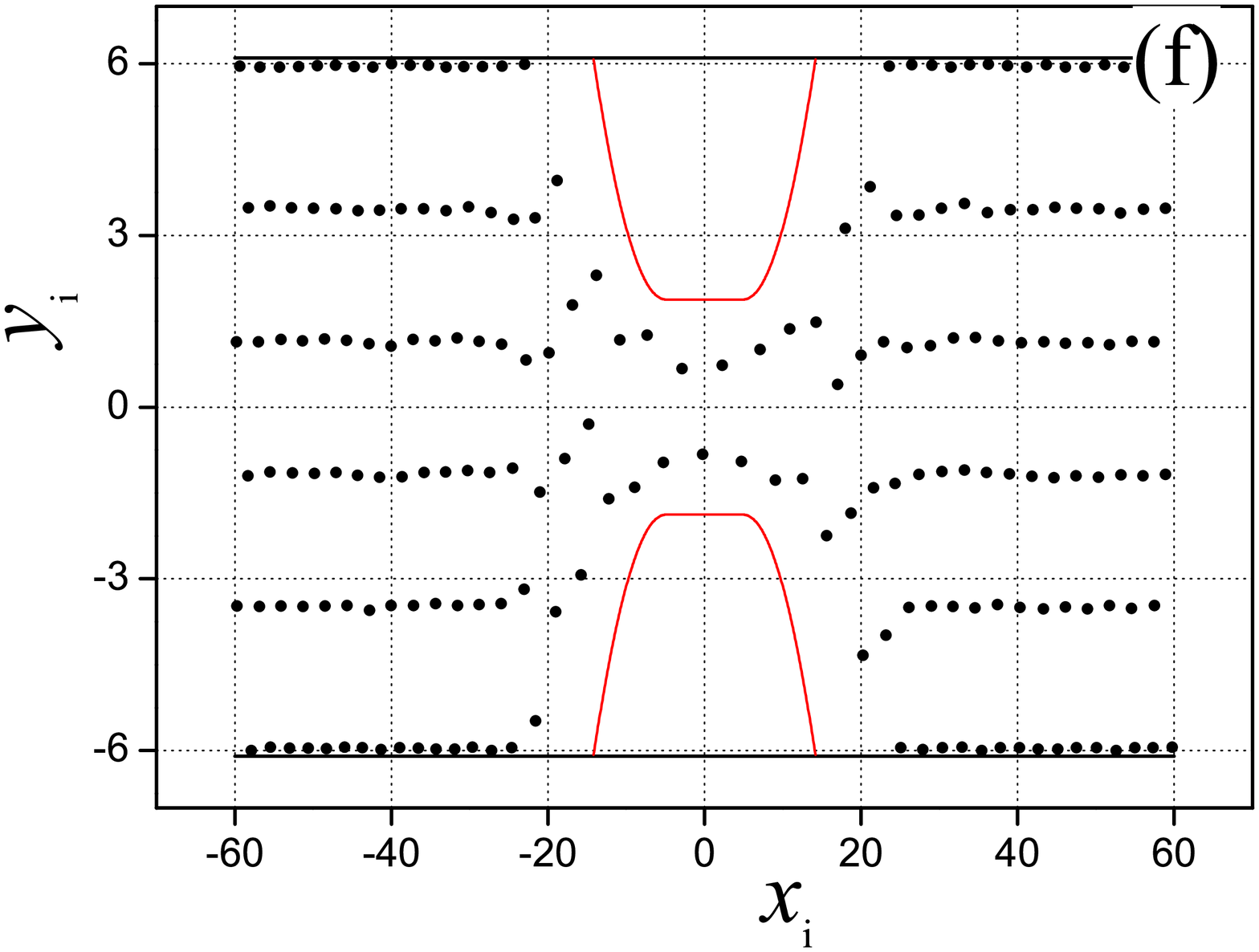,width=9cm}}
\vspace*{-0.4cm}
\centerline{\epsfig{figure=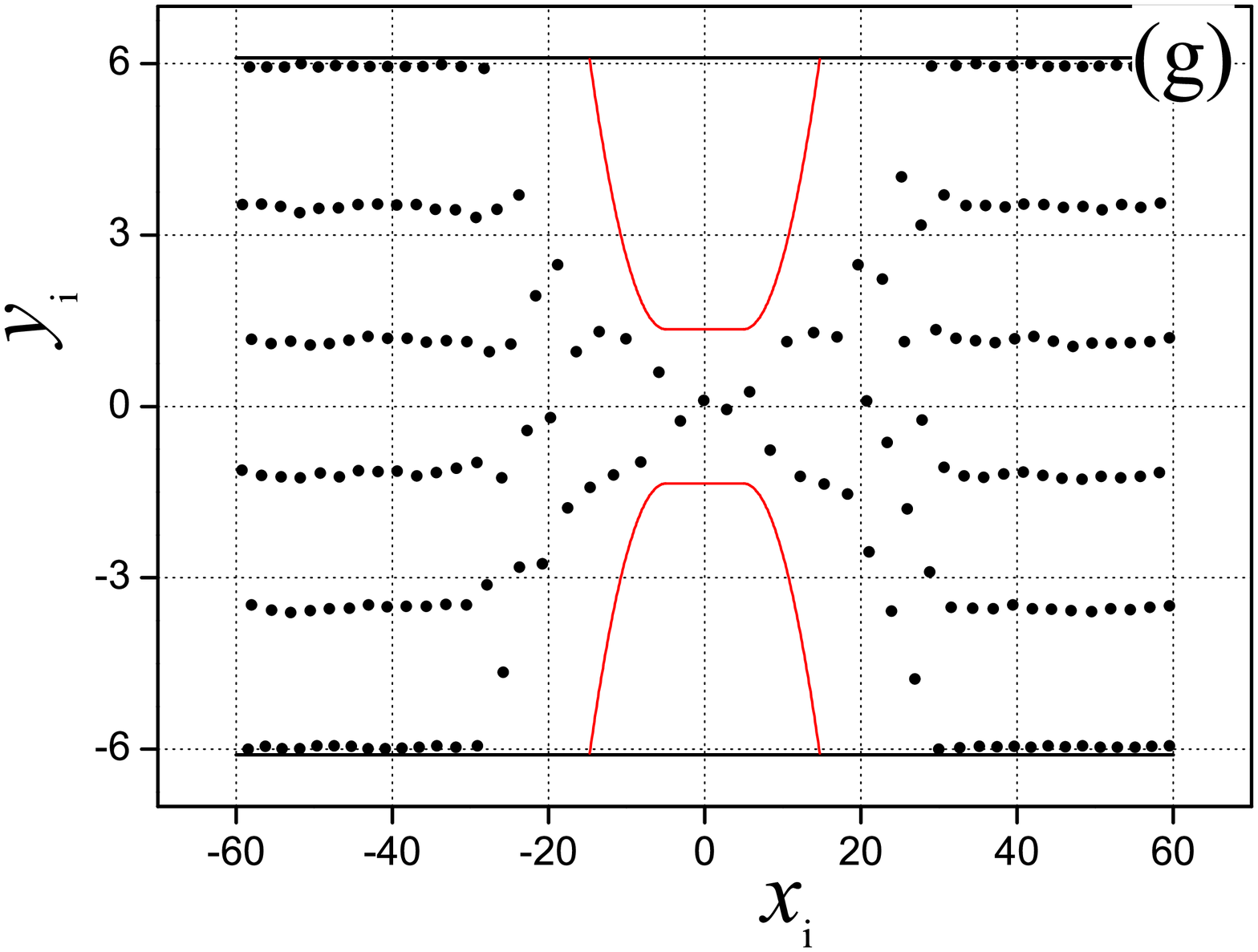,width=9cm}\hspace*{-1.2cm}\epsfig{figure=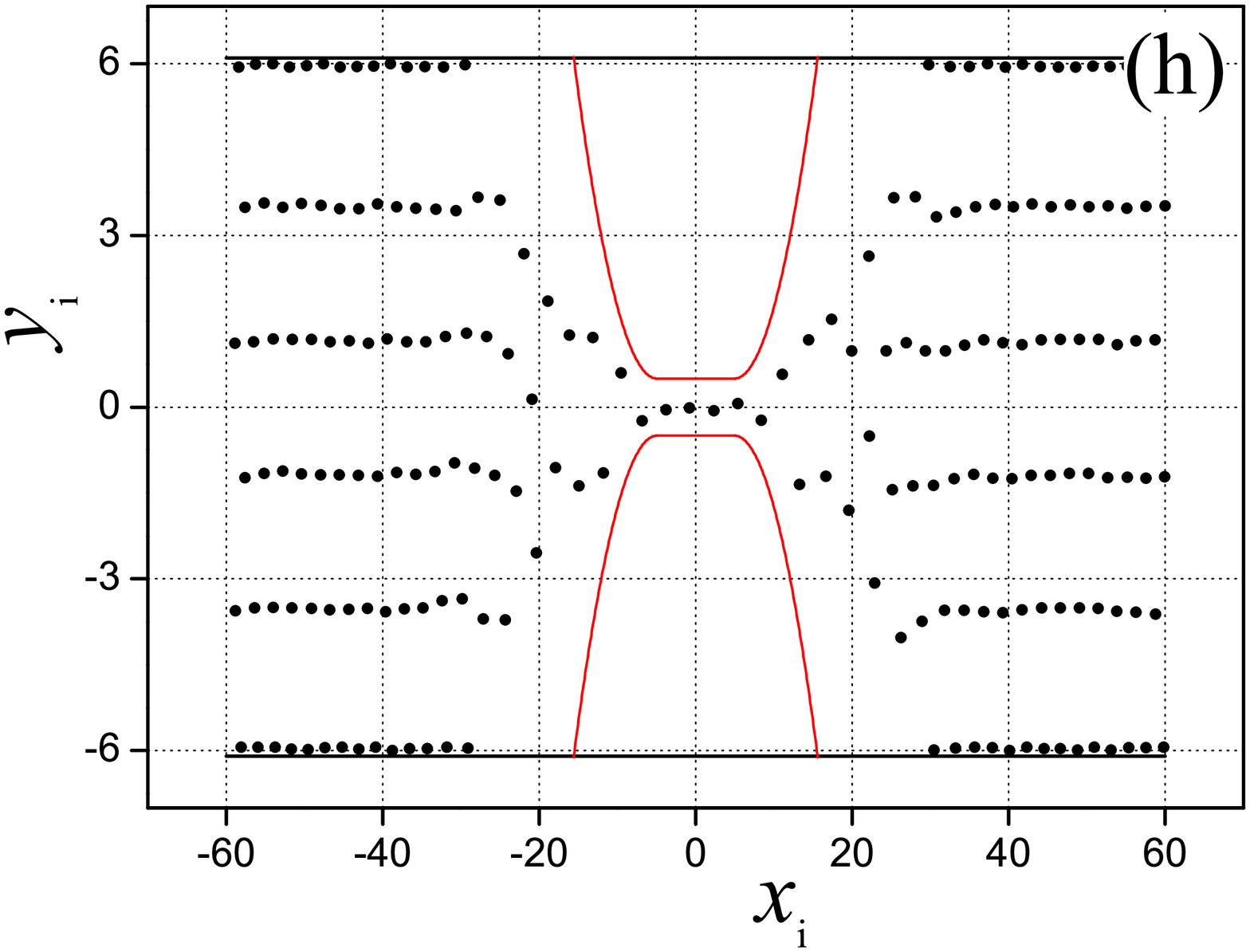,width=9cm}}
\vspace*{-0.4cm}
\caption{Electron distributions for the system with modified constriction [Eq.(\ref{SPPlatoPot})], for different values of the potential parameter $V_{g}$ : (a) 0.2, (b) 0.4, (c) 0.57, (d) 0.6, (e) 0.8,  (f) 1.2, (g) 1.35, (h) 1.6; 200 particles, $f_{x}=0.015$.}
\label{SnShL4F015}
\end{figure*}

Using this modified saddle-point-like potential results in more pronounced different regimes, as shown in 
Figs.~\ref{BifDL4Lp4F015} and \ref{SnShL4F015}. 
It is worth noting that in this phase diagram 
(i.e., the number of rows versus $V_{g}$) 
we observe the transition ``1-2-4-3-6-4-5'' 
with the striking inversions ``2-4-3'' and ``3-6-4'' 
(cp. to Refs.~\cite{Piacente2004,Piacente2005}). 
These transitions have not been observed (nor found in calculations) in similar classical systems, e.g., colloids moving in narrow channels or vortices in superconducting narrow stripes. 

\begin{figure}[t!]
\vspace*{-6mm}
\centerline{\hspace*{-1mm}\epsfig{figure=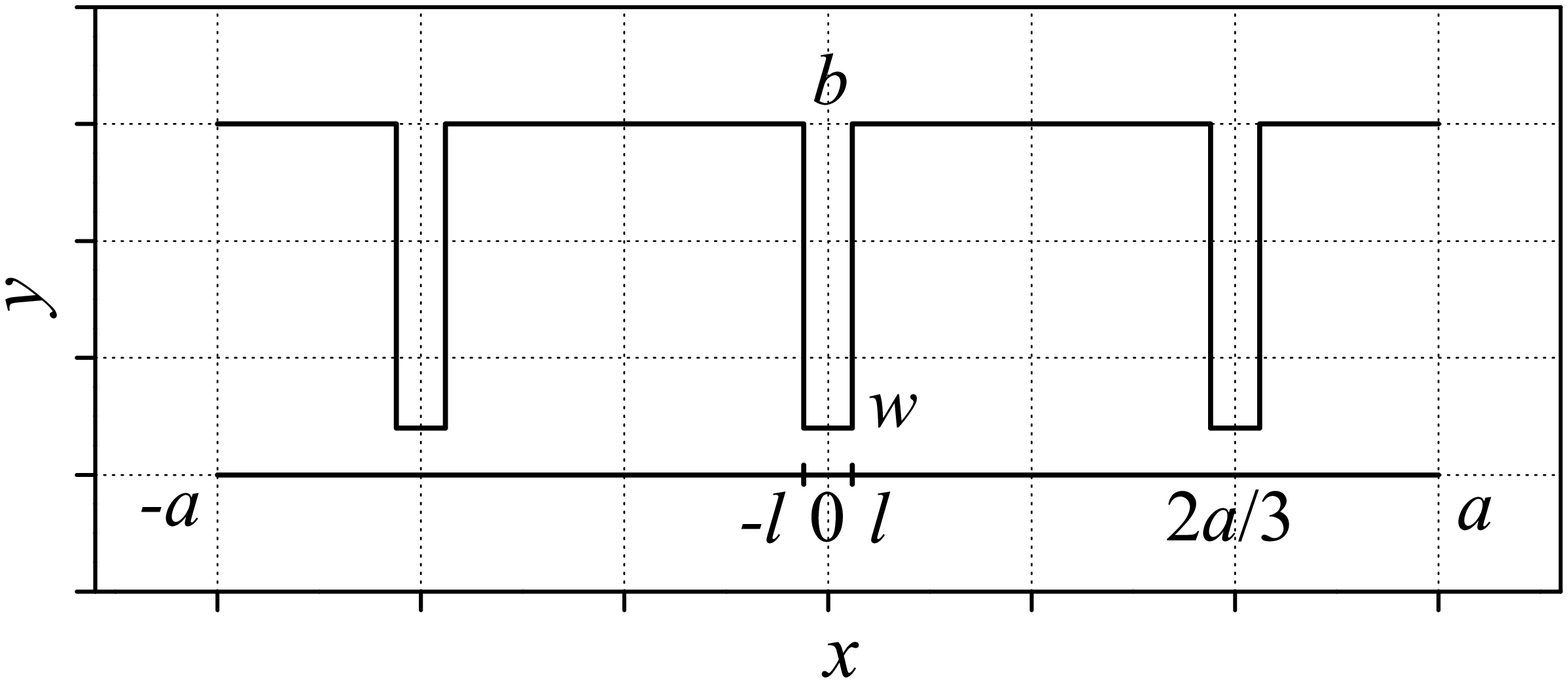,width=9.6cm}}
\vspace*{-0.5cm}
\caption{Sketch of the channel with three asymmetric constrictions, the lateral walls are vertical with the angle 90$^\circ$.}
\label{Chan3ConSketch} 
\end{figure}
\begin{figure}[t!] \vspace{-0.6cm} 
\centerline{\hspace*{0.8cm}\epsfig{figure=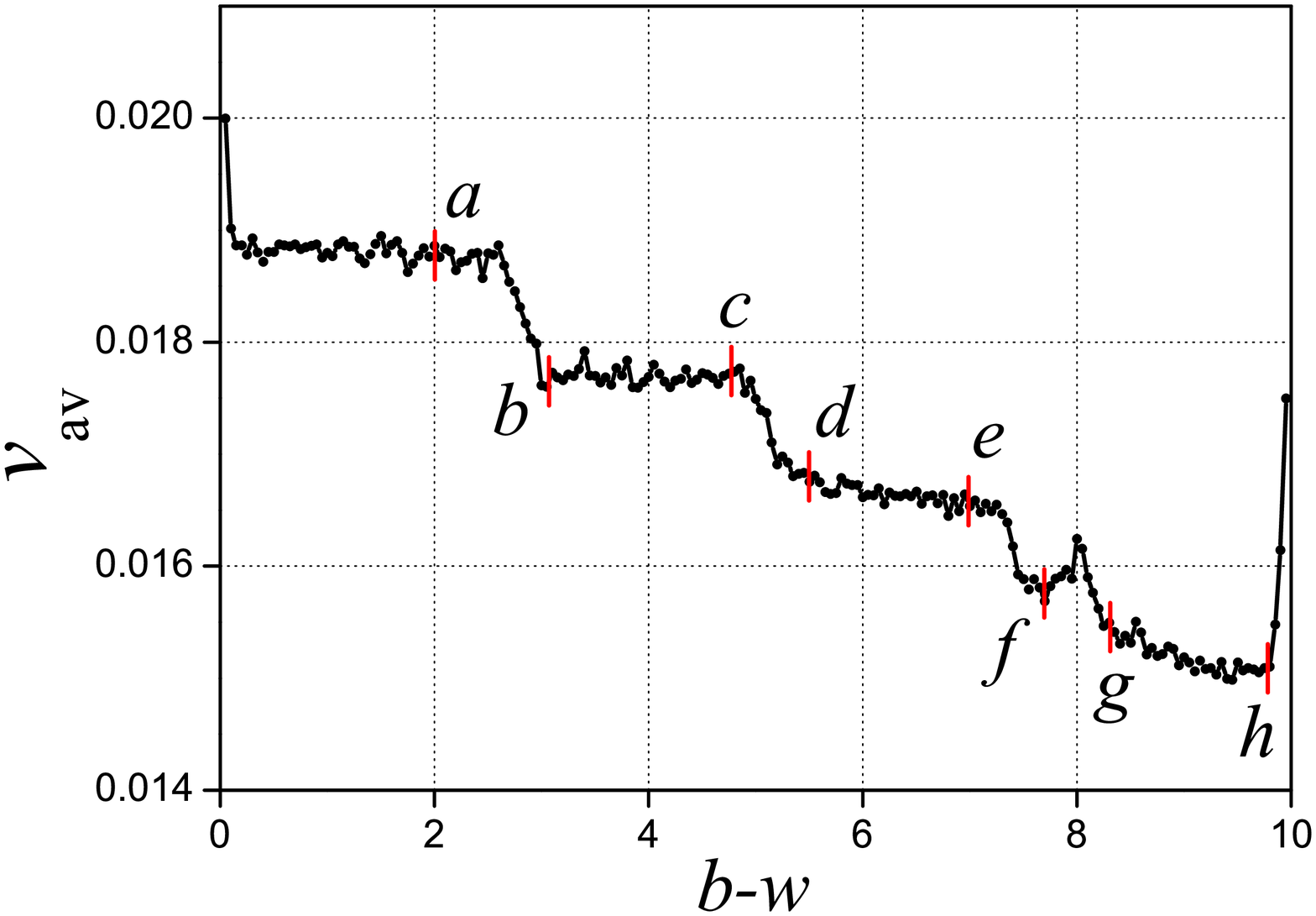,width=9cm}  }
\vspace*{-0.3cm}
\caption{Graphs of the average velocity of all the electrons in the channel in the $x$-direction; 300 particles, $f_{x}=0.02$, $f_{y}=0.0$.} 
\label{3ConStairsW}
\end{figure}

\section{Channel with asymmetric constrictions} 

Let us consider now a channel with asymmetric constrictions. In our model, three constrictions are placed such that they divide the channel into three equal in length compartments, as shown in the sketch in Fig.~\ref{Chan3ConSketch}. 
The lateral walls of the constrictions are vertical, and all three constrictions have the same width that is denoted with $w$. 
The interaction of the particles with the boundaries of the channel and the constrictions is hard-wall. 
As before, the longitudinal driving force $f_{x}$ is applied, and we also introduce the additional {\it transversal} force in the $y$-direction $f_{y}$. 
In this structure with geomteric constrictions, we vary the width of the constriction $w$ and we study how this change influences the transport of the particles. 

\begin{figure*}[t] %%\vspace{-0.6cm} 
\centerline{\epsfig{figure=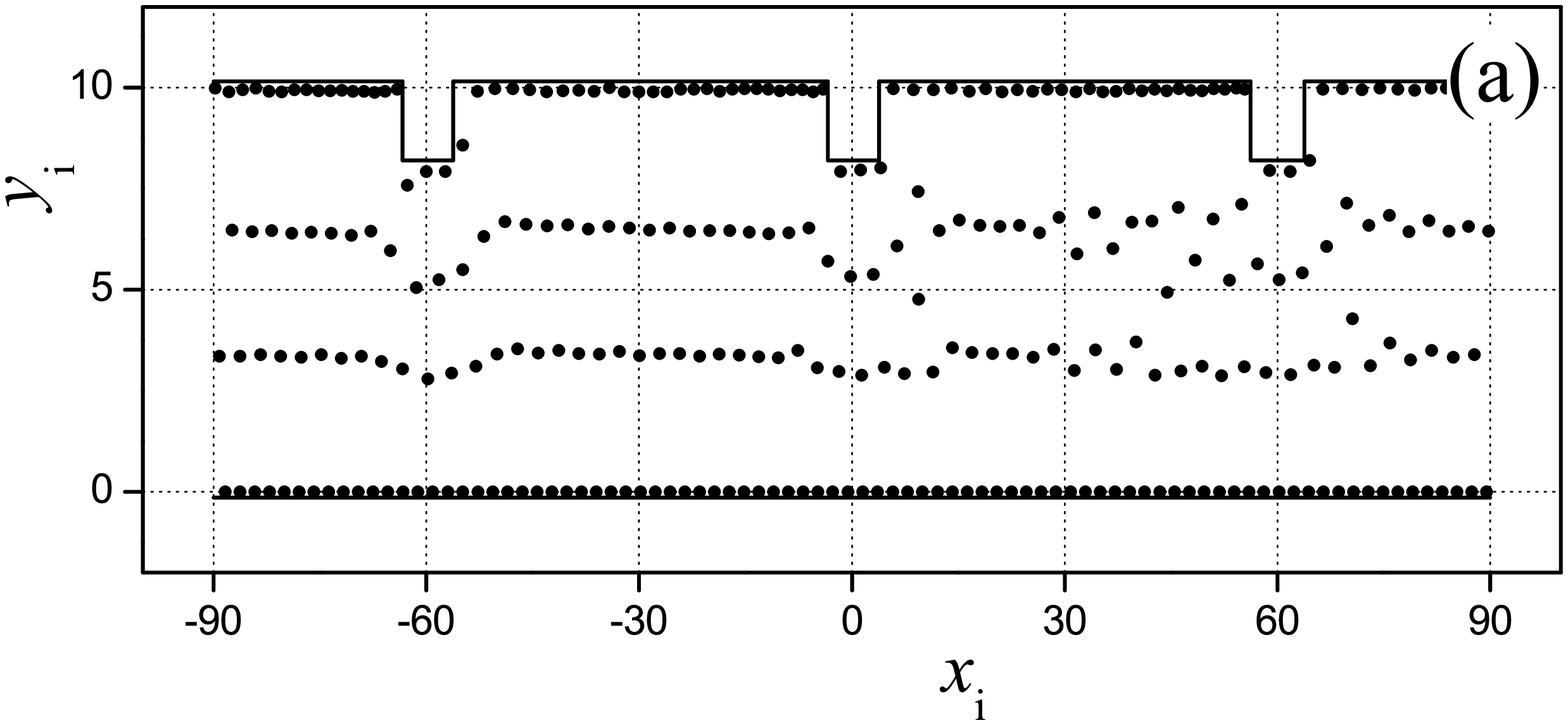,width=9cm} \hspace{-0.8cm} 
\epsfig{figure=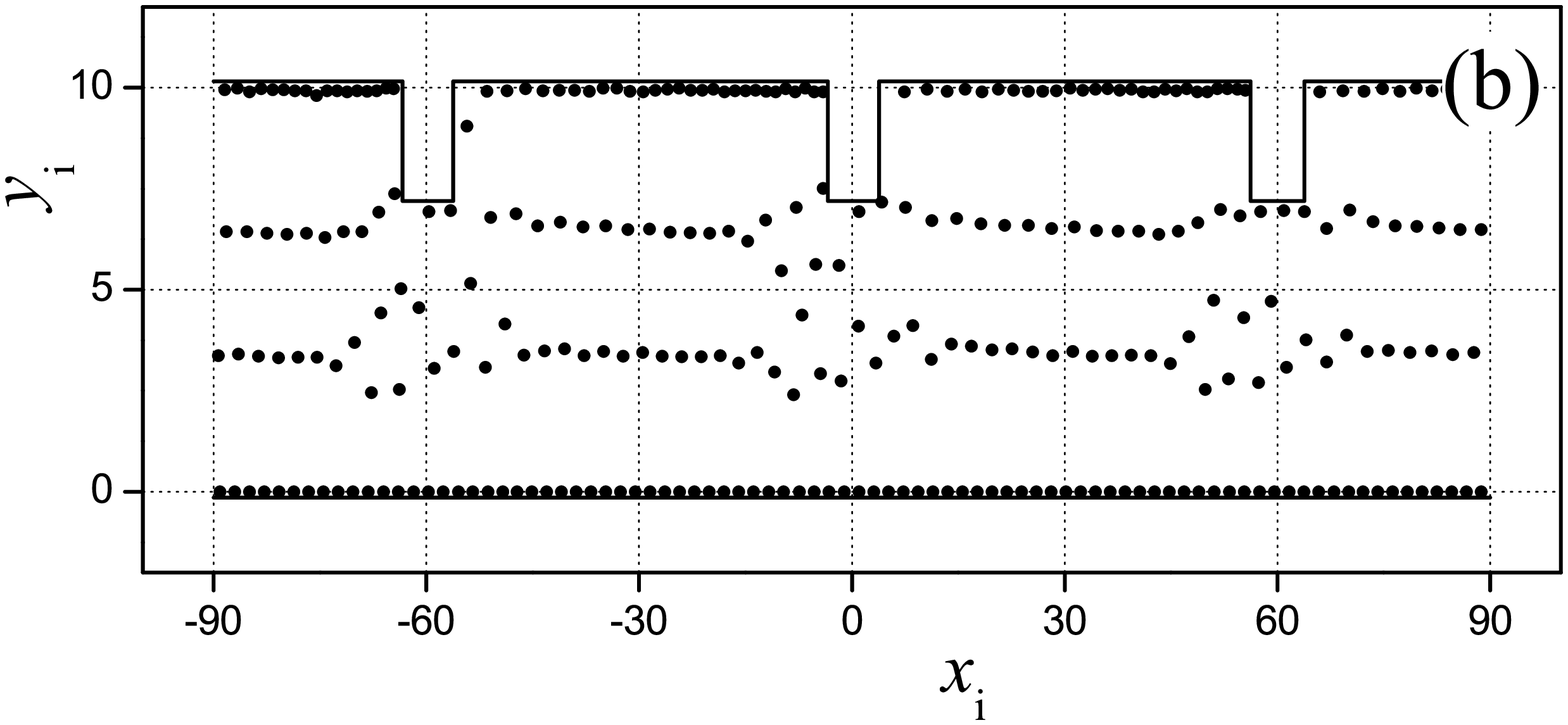,width=9cm} }
\vspace*{-0.4cm}
\centerline{\epsfig{figure=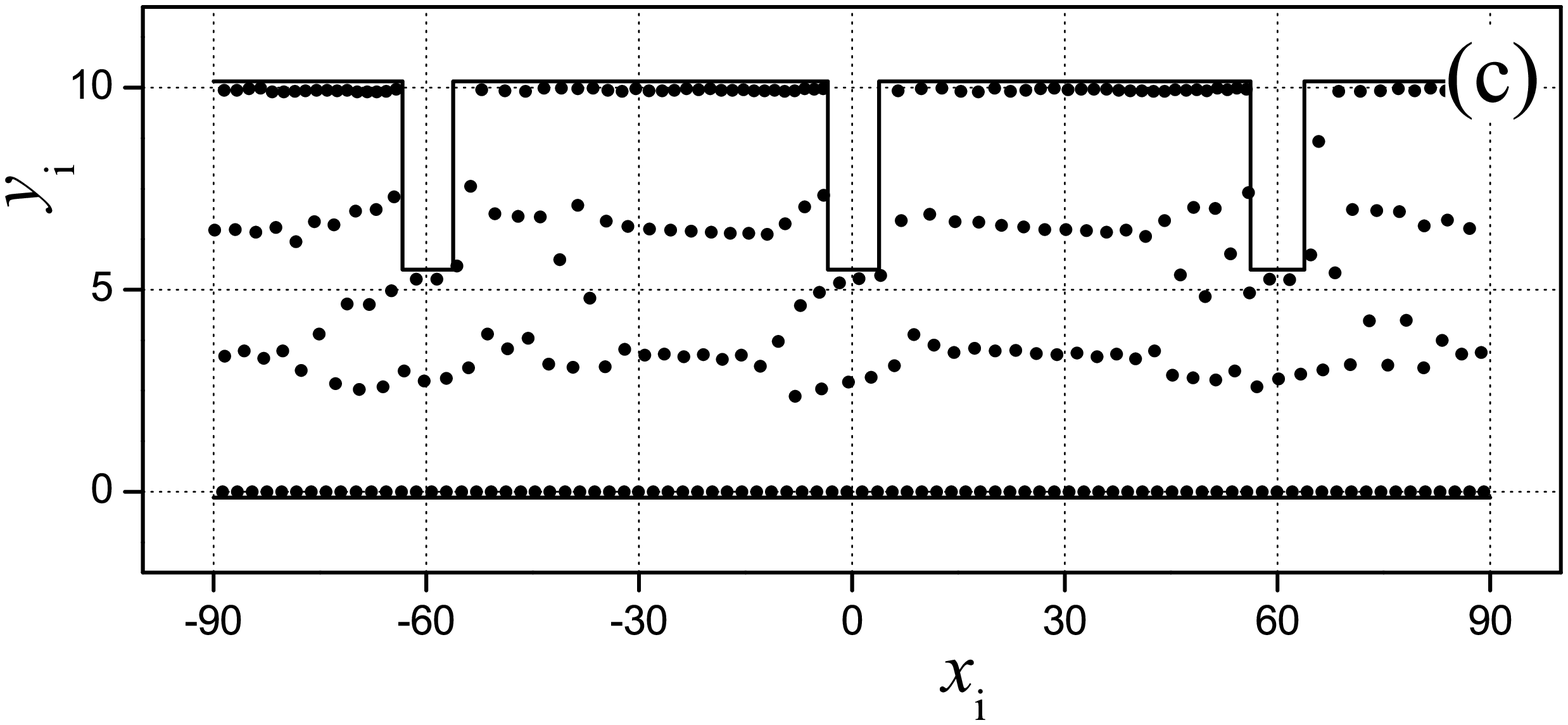,width=9cm} \hspace{-0.8cm} 
\epsfig{figure=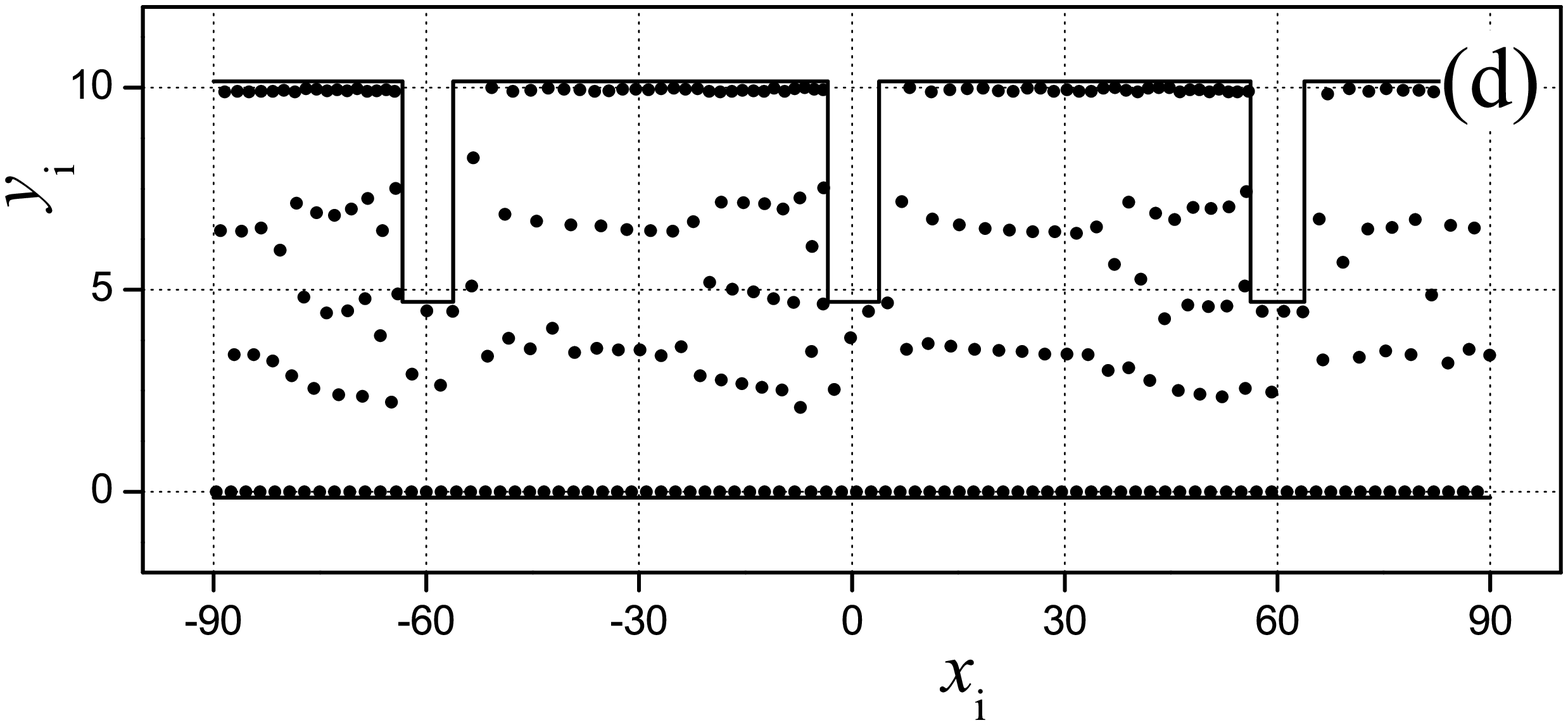,width=9cm} }  
\vspace*{-0.4cm}
\centerline{\epsfig{figure=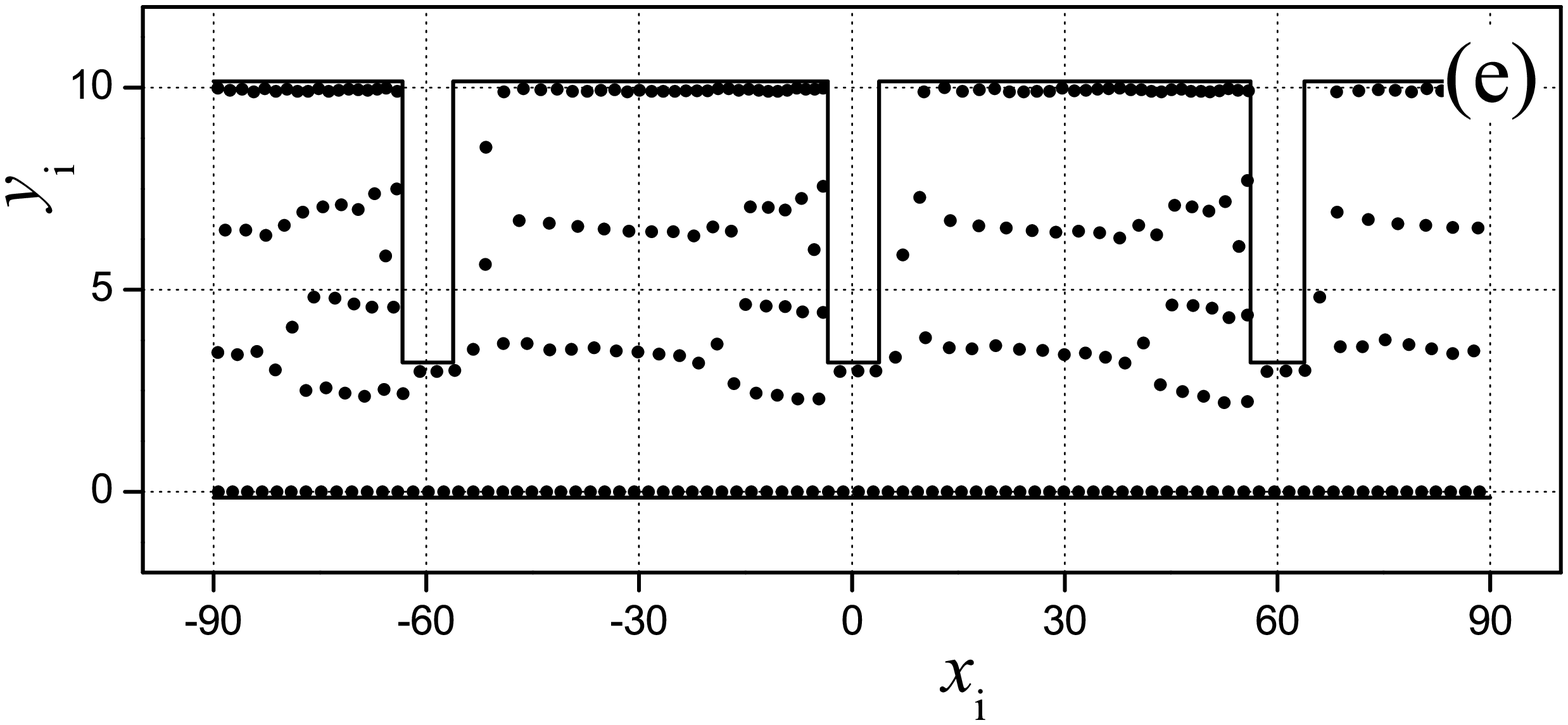,width=9cm} \hspace{-0.8cm} 
\epsfig{figure=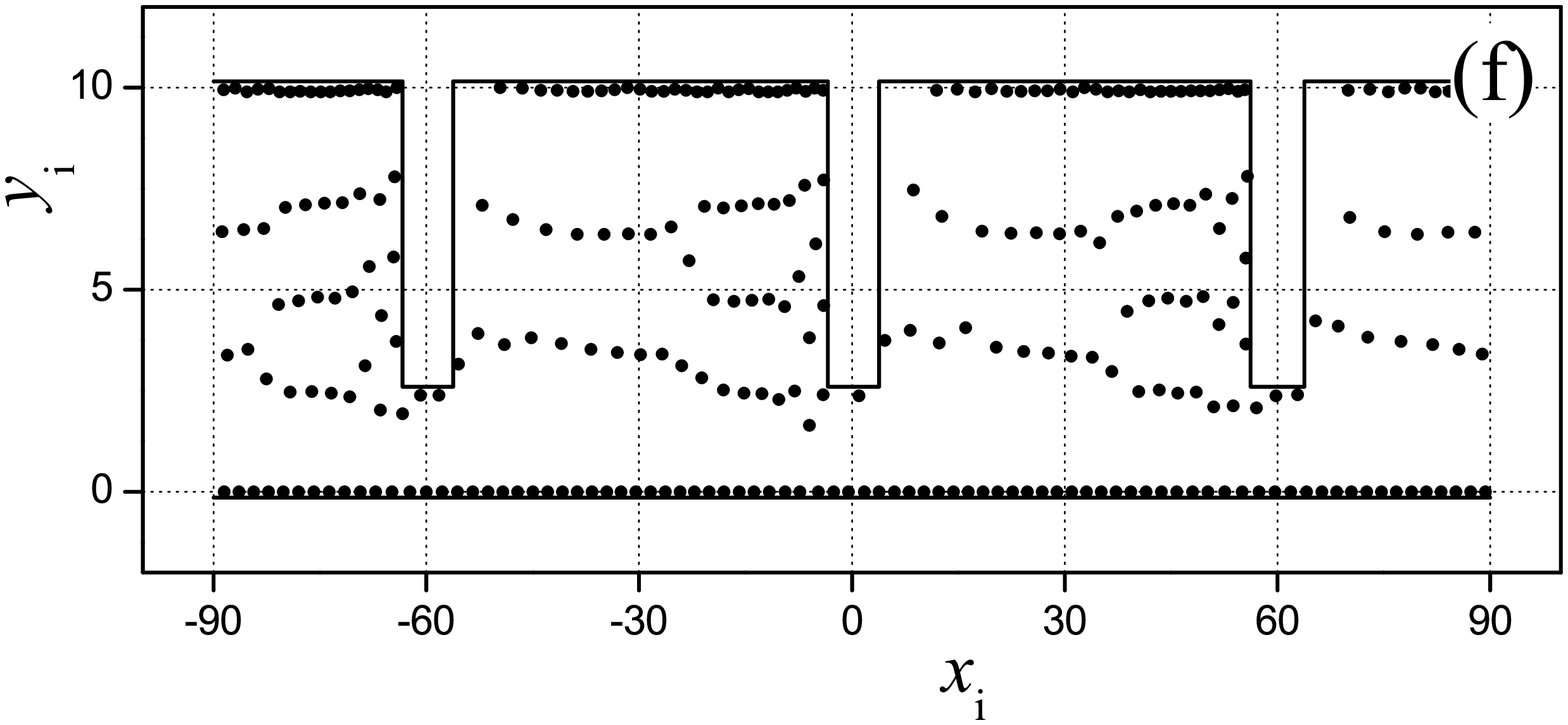,width=9cm} }
\vspace*{-0.4cm}
\centerline{\epsfig{figure=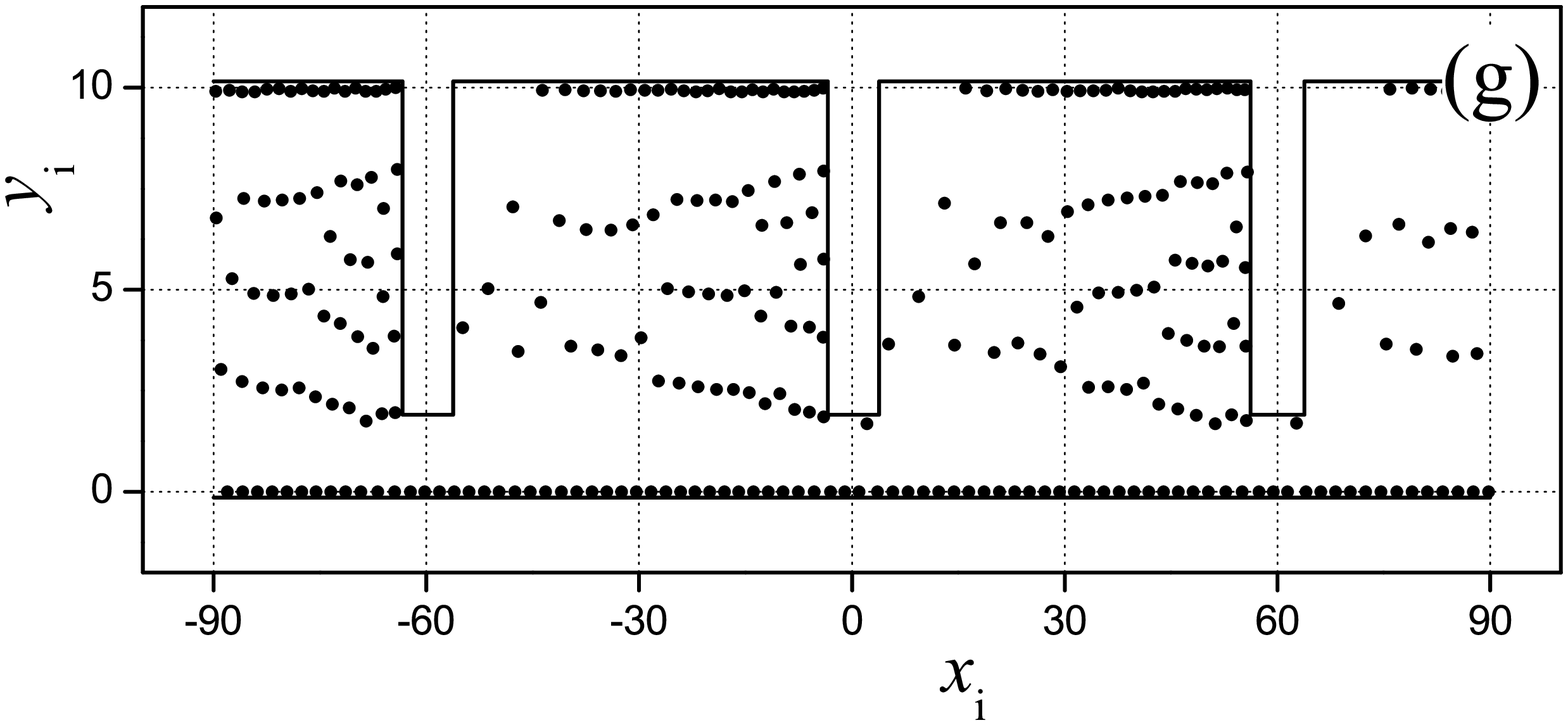,width=9cm} \hspace{-0.8cm} 
\epsfig{figure=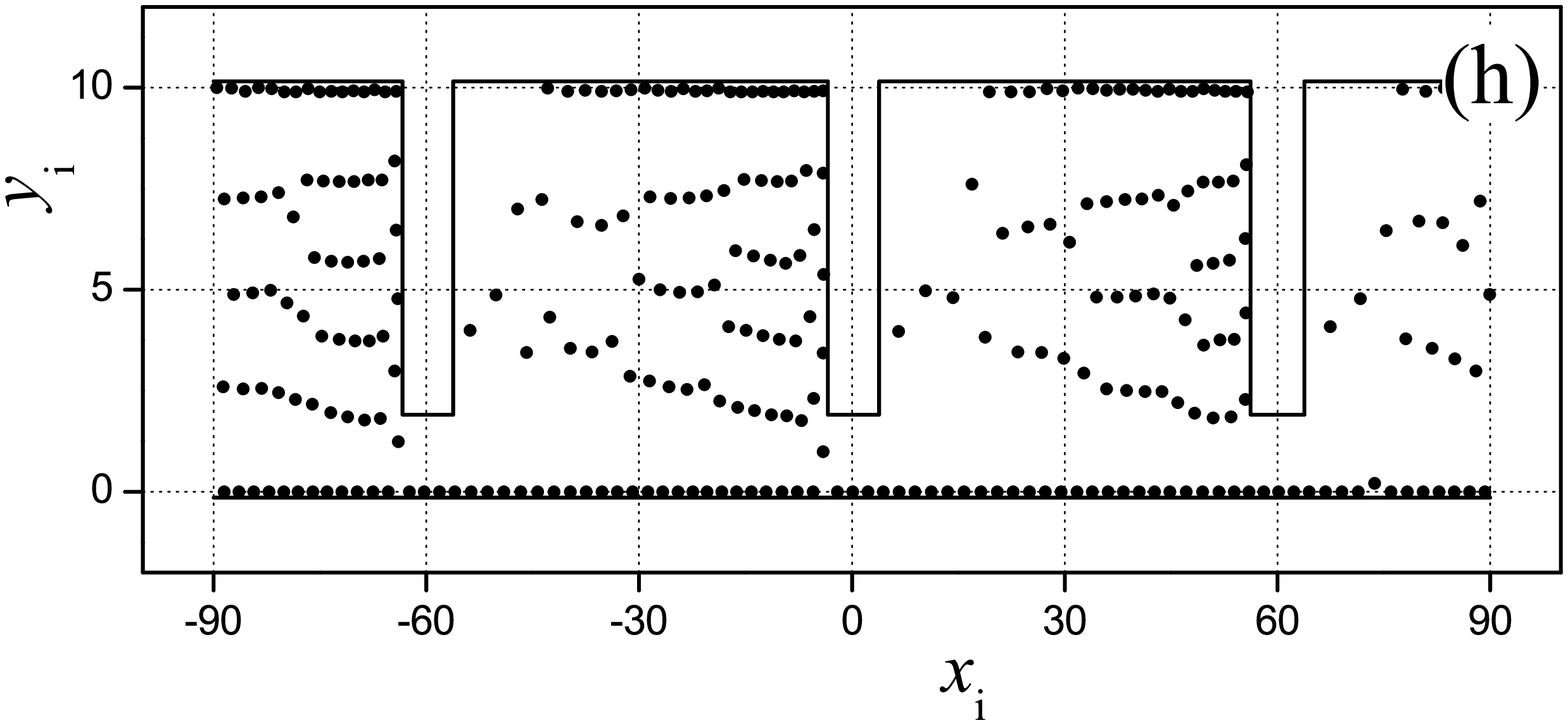,width=9cm} }
\vspace*{-0.4cm}
\caption{Electrons distributions for different values of the constriction parameter $b-w$: (a) 2,
 (b) 3, (c) 4.7, (d) 5.5, (e) 7,  (f) 7.6, (g) 8.3, (h) 9.8; 300 particles, $f_{x}=0.02$, $f_{y}=0.0$. }
\label{Examples3Con}
\end{figure*}

The resulting average velocity of the particles in the channel as a function of the gap $b-w$ exhibits stair-like structure, as shown in Fig.~\ref{3ConStairsW}. 
Here, we apply a constant weak driving force $f_{x}=0.02$.  The characters (a to i) mark the parameter values for the corresponding electrons distributions shown in 
Fig.~\ref{Examples3Con}. 

The electron distributions shown in Fig.~\ref{Examples3Con} allow us to understand the dynamics and to explain the appearance of the stairs in the average velocity curve. 
Thus each transition to the lower stair (e.g., a to b, c to d, etc.) in principle occurs when the constriction gap $b-w$ decreases by the value corresponding to one spacing between the adjacent rows in the particle distribution. 
In practice, however, the situation appears to be more complicated since the perfectly ordered row structure (observed for a channel without constrictions) is strongly influenced by the asymmetric constrictions. 
For example, as one can see in Fig.~\ref{Examples3Con}(a), even in case of a wide gap $b-w$ the four-row structure (including the rows of particles near the boundaries) is modified near the constrictions. 
The row near the constriction is locally split into two, and the next row is cuvred. 
As a result, a significant drop in the $v_{av}$ is observed, although the gap $b-w$ is still wide enough to allow a free motion of the {\it undisturbed} row of particles (i.e., in the absence of the constrictions). 
Therefore, the reason for the observed drop in the $v_{av}$ is that the row of particles splits near the constriction, and the particles of one branch of the splitted area become  blocked by the constriction (see Fig.~\ref{Examples3Con}(b)). 
As a consequence, some additional decrease of the gap $b-w$ (such that the geometrical line connecting particles in the row away from the constrcition crosses the constriction) does {\it not} result in any decrease in the $v_{av}$ (see Fig.~\ref{Examples3Con}(c) and Fig.~\ref{3ConStairsW}). 
In a similar manner, other stairs in the function $V_{av}(b-w)$
shown in Fig.~\ref{3ConStairsW} are explained by analyzing the particle distributions presented in Fig.~\ref{Examples3Con}. 

\begin{figure}[t!] \vspace{-0.6cm} 
\centerline{ \epsfig{figure=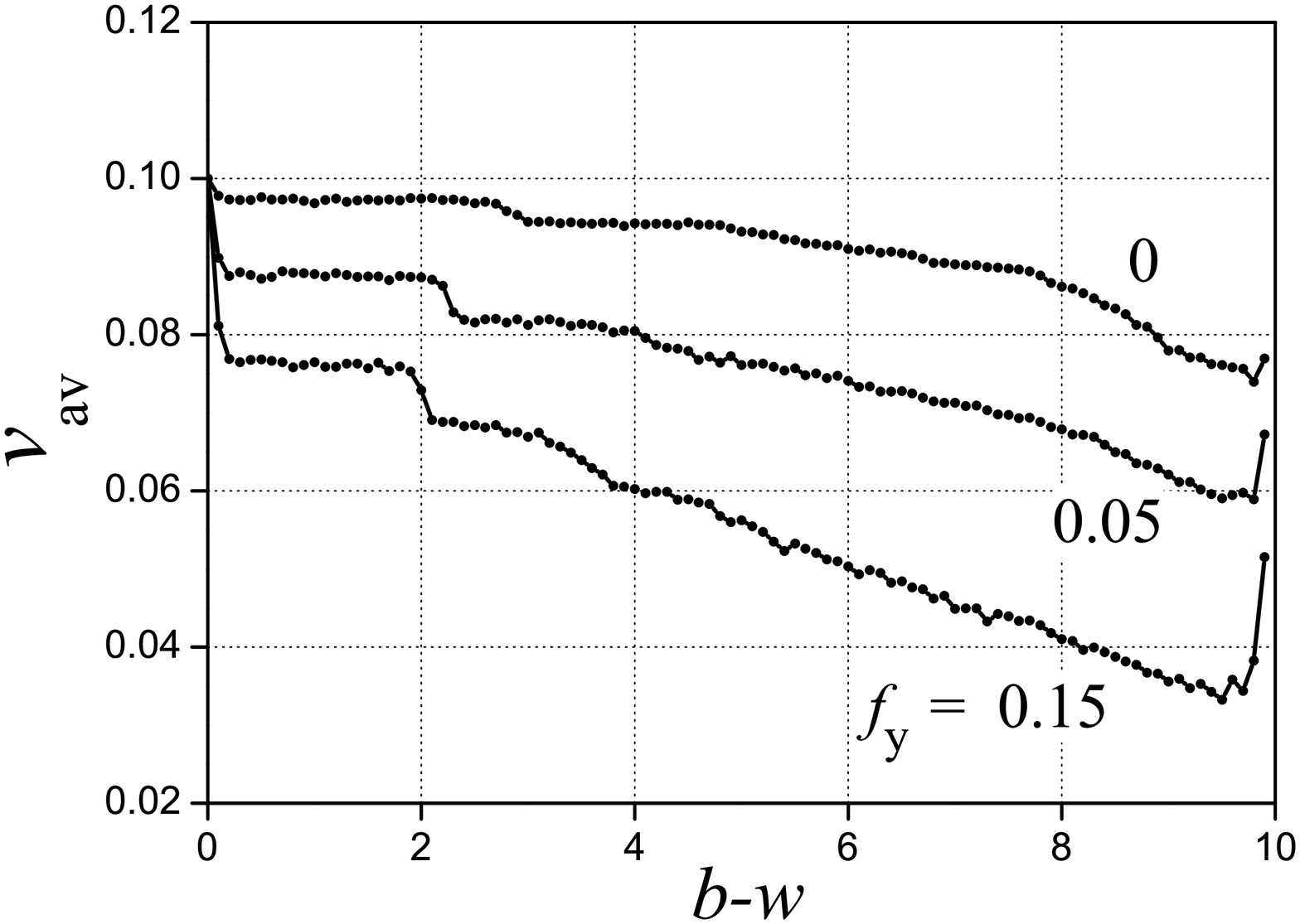,width=9cm} }
\vspace*{-0.3cm}
\caption{Graphs of the average velocity of all the electrons in the channel in the $x$-direction; 200 particles, $f_{x}=0.1$.} 
\label{3ConStairsWdifF}
\end{figure}

As we argued above, it is important to apply a very weak driving force that provides a quasi-stationary transport regime when the particle distribution is characterized by a row structure. 
As we just showed, this is the row structure that is responsible for the stair-like $V_{av}(b-w)$-curve. 
To check this statement, we applied slightly stronger driving: $f_{x}=0.1$ and we found that the stairs become smoothened as shown in Fig.~\ref{3ConStairsWdifF}. 
On the other hand, we can control the depth of the steps by applying additional transversal force $f_{y}$. 
Let us discuss this effect. 

\begin{figure*}[t!] \vspace{-0.6cm} 
\centerline{\epsfig{figure=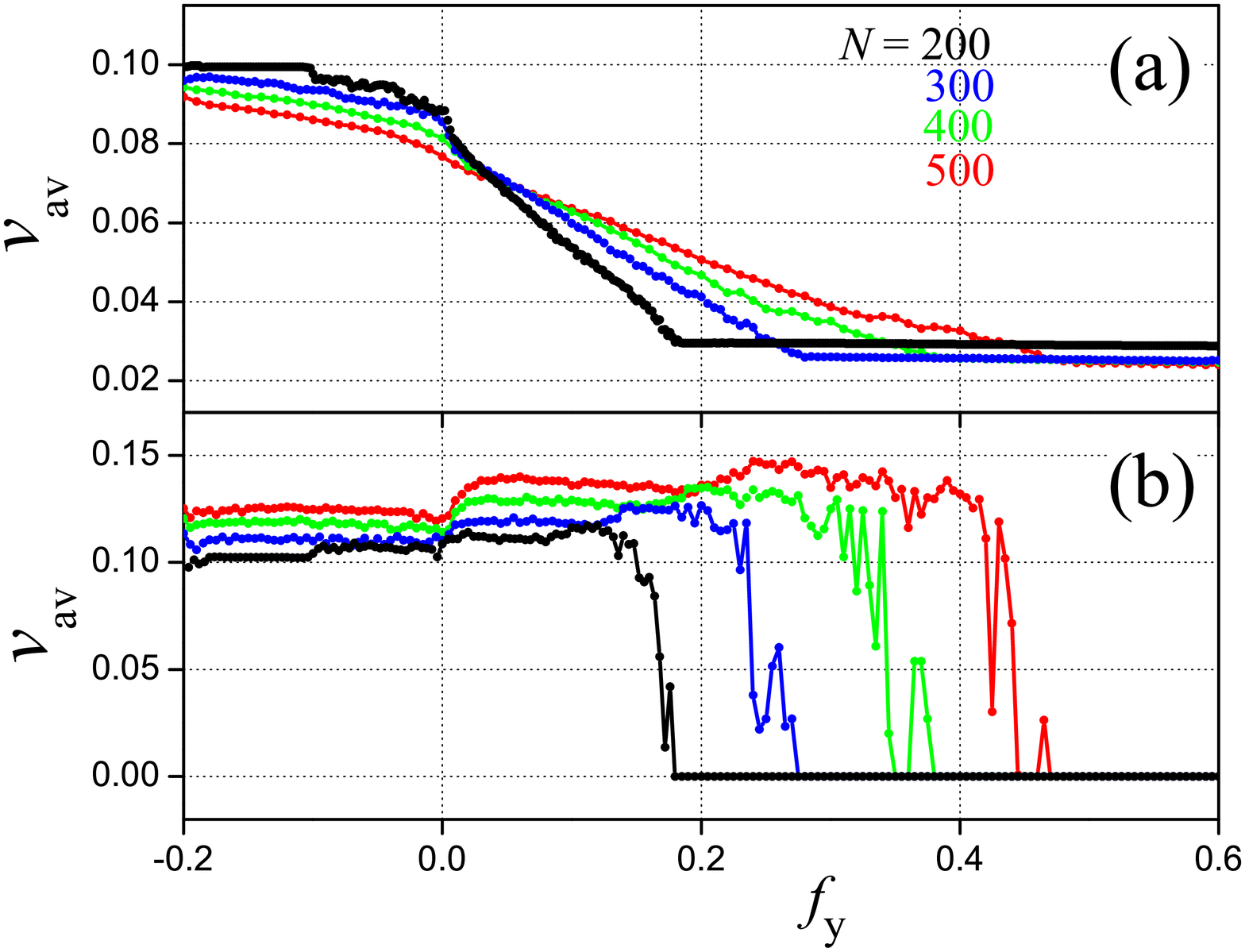,width=9cm} \hspace{-0.5cm} 
\epsfig{figure=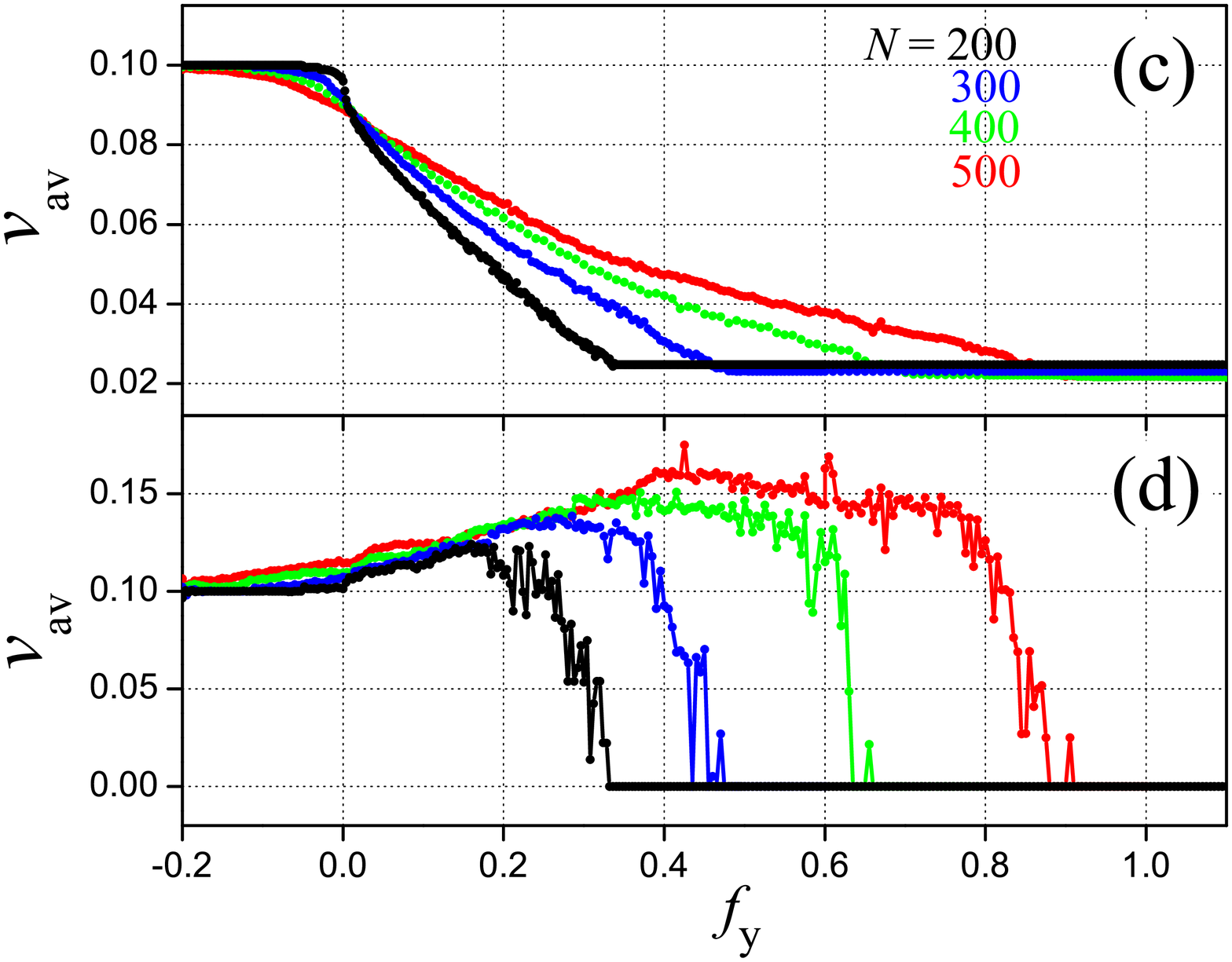,width=9cm} }
\caption{Graphs of the average velocities of all the electrons in the channel (a, c) and electrons in the central constriction near $x = 0$ (b, d) in the $x$-direction, $f_{x}=0.1$, $w$ = 2 (a, b) and 5 (c, d). }
\label{3ConDifW}
\end{figure*}

We fix the constriction width $w$ and the longitudinal force $f_{x}$, and investigate the effect of varying the transversal force $f_{y}$ on the average velocity of the particles $V_{av}$. 
We apply a relatively large value of the driving force $f_{x}=0.1$~\cite{strong}, and in this case the average velocity of all the particles in the channel does not exhibit pronounced stairs, but this effect can be enhanced if we consider the average velocity of the particles in the constriction, as shown in Fig.~\ref{3ConDifW}. 
In this figure, 
we plotted the average velocity for different values of the total number of particles in the channel $N=200$, 300, 400, and 500, 
and we found that the stairs are more pronounced for the case of the narrow constriction with $w=2$, Fig.~\ref{3ConDifW}(a, b) rather than for a wide constriction. 

The important result is that we can control the particle flow through the channel with asymmetric constrictions not only by changing the size of the constriction but, more strikingly, by tuning the transversal force $f_{y}$ applied to the channel (for a fixed constriction size). 
The latter is an analog of electronic FET-devices and it can be easily realized in experiment. 

\begin{figure*} %%\vspace{-0.6cm} 
\centerline{\epsfig{figure=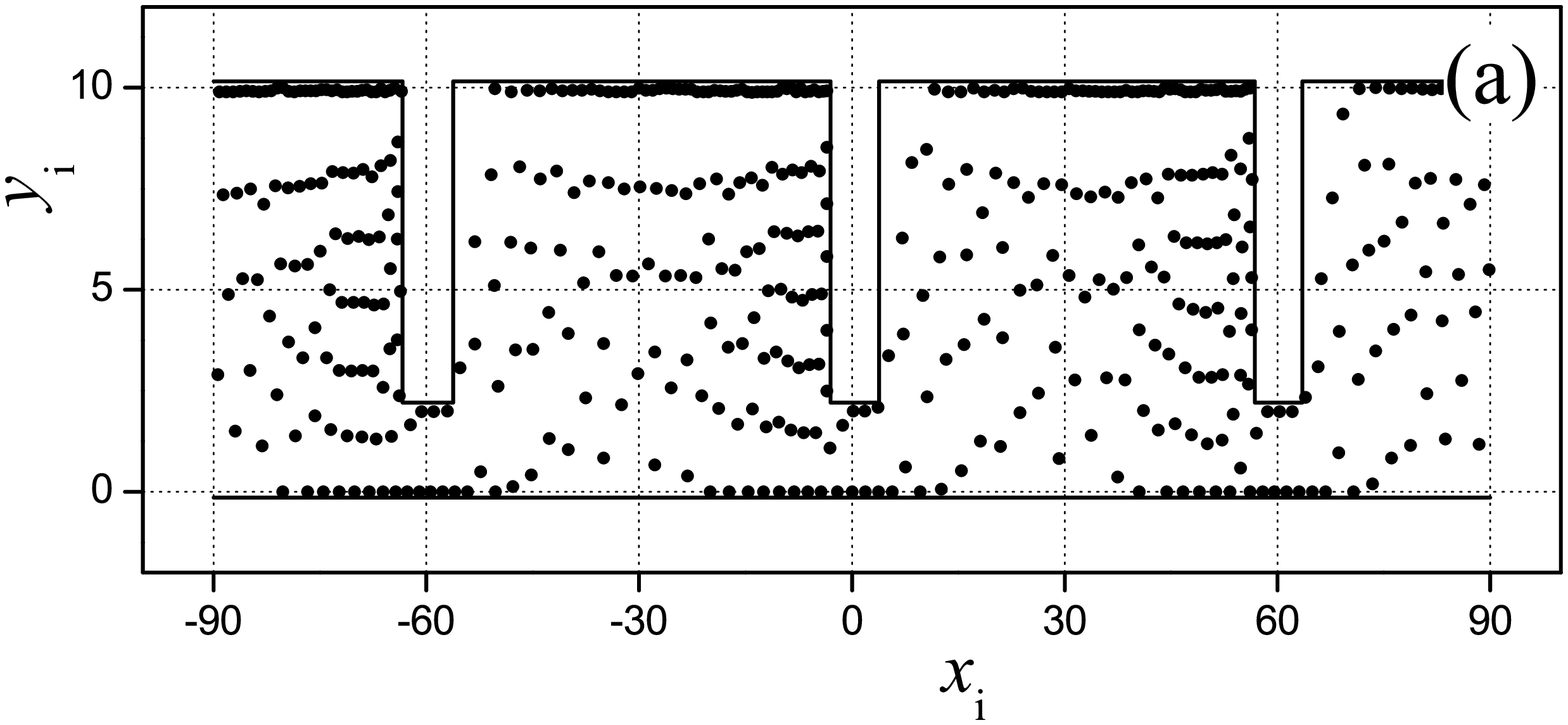,width=9cm} \hspace{-0.8cm} 
\epsfig{figure=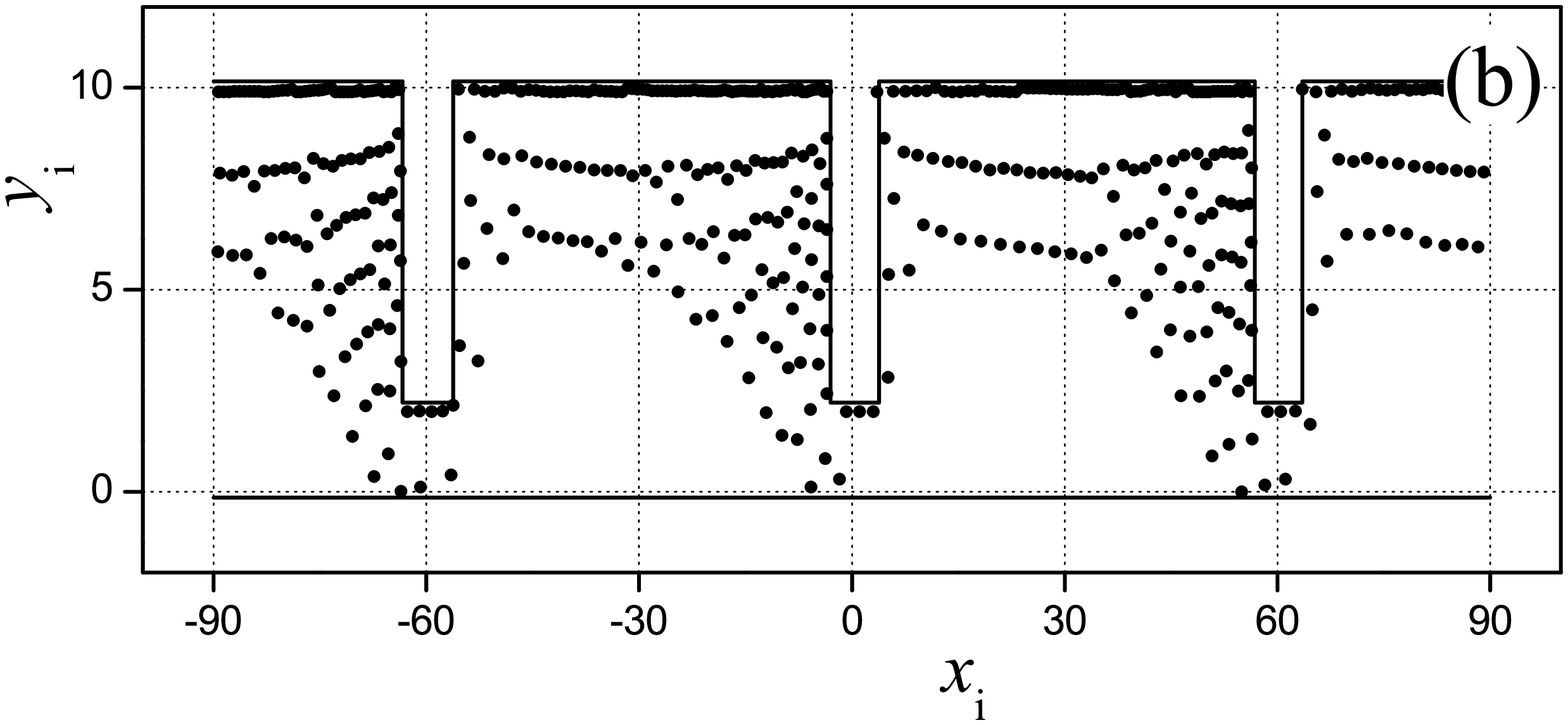,width=9cm} }
\vspace*{-0.4cm}
\centerline{\epsfig{figure=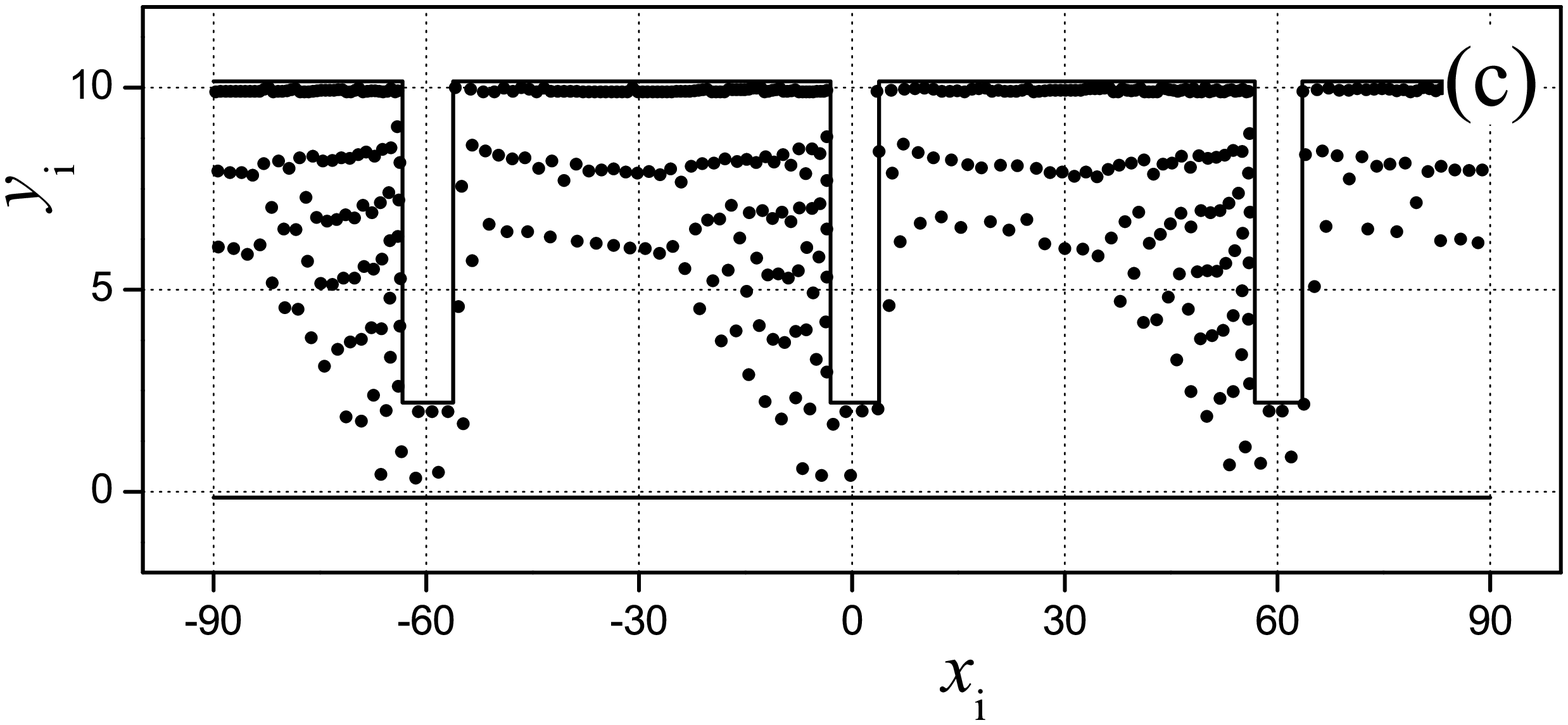,width=9cm} \hspace{-0.8cm} 
\epsfig{figure=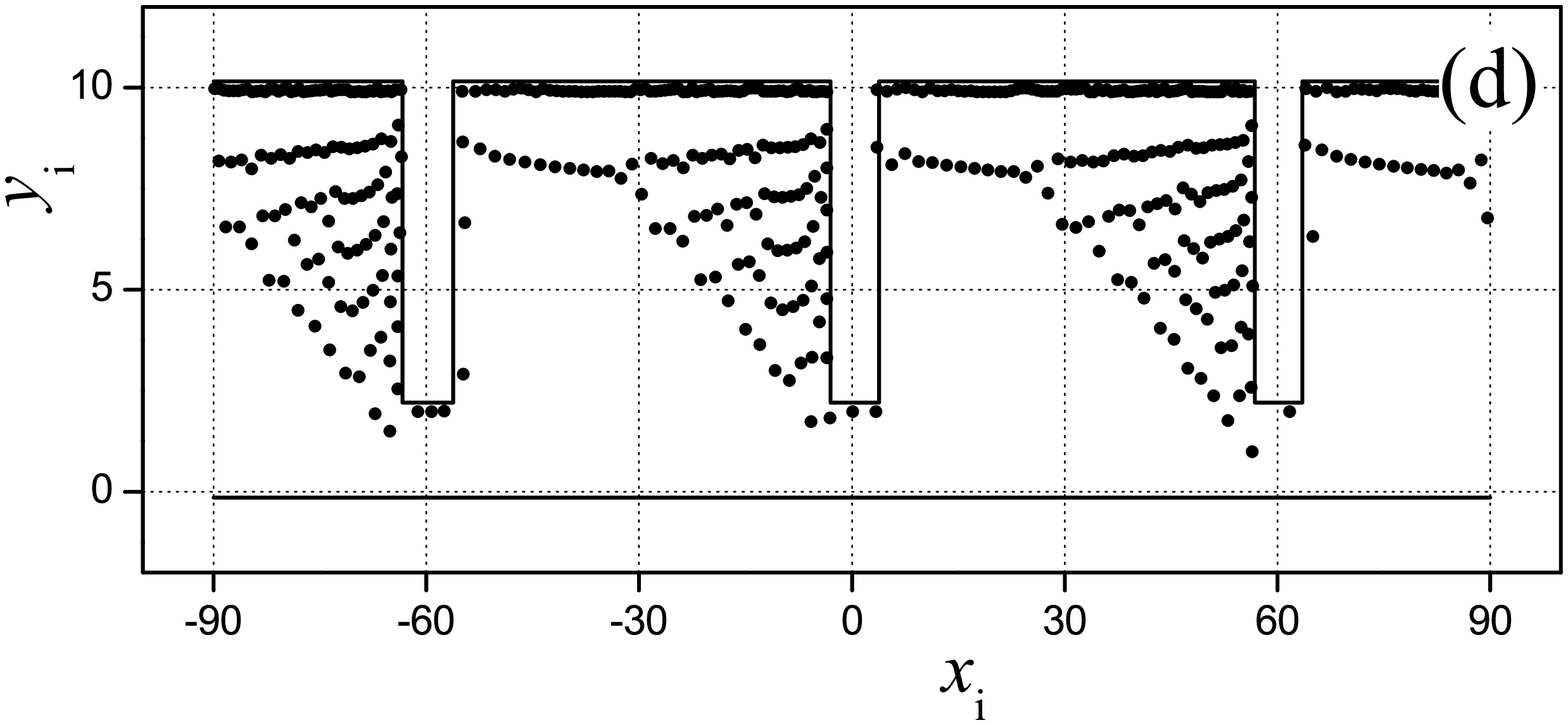,width=9cm} }  
\vspace*{-0.4cm}
\caption{Electrons distributions for different values of the transversal force $f_{y}$: (a) 0.05,
 (b) 0.2, (c) 0.23, (d) 0.35; 500 particles, $w = 2$, $f_{x}=0.1$. }
\label{Examples3ConFy}
\end{figure*}

Finally, Fig.~\ref{Examples3ConFy} shows the particle distributions corresponding to the above case of $N=500$  particles shown in Fig.~\ref{3ConDifW}(b). 
The distributions indicate that the electron distributions is rather far from the quasi-equilibrium, it is characterized by pronounced gradients in the particle density. 
Nevertheless, the calculated distributions reveal local row structure which is responsible for the appearance of the steps in the $v_{av}(f_{y})$-curve. 


\section{Conclusions} 

By numerically solving the Langevin equation of motion for a system of interacting charged particles, we investigated the transport properties of the Wigner crystal driven by an external force on the surface of a superfluid $^{4}$He 
in narrow channels with constrictions. 
The width of the constriction varied such that it could accommodate from a few rows of electrons 
(i.e., the quasi-one-dimensional regime) 
to ultimately just one row, i.e., the ``quantum wire'' regime, when the channel width is comparable to the inter-electron separation. 
We analyzed the average velocity of the moving electrons through the constrictions as a function of the parameters: the strength of the driving force, the shape of the constriction (i.e., with either inclined or normal boundaries with respect to the channel walls) and the strength of the model gate voltage. 

We considered two different types of constrictions: symmetric with respect to the central axis of the channel, and asymmetric constrictions. 

In case of symmetric constrictions, we addressed a recent observation revealed in the experiments on measuring the conductance of a classical point contact, for long and short constrictions. 
In particular, in our simulations we revealed a significant difference in the dynamical behavior for long and short constrictions. 
Namely, the oscillations of the average velocity of particles in case of short constrictions exhibited a clear correlation with the transitions between the states with different numbers of rows of particles in the constriction, while for the systems with longer constriction these oscillations are suppressed. 
The obtained results are in agreement with the recent experimental observations. 

We also proposed to use a narrow channel with asymmetric constrictions as a FET-like device for an effective control of the electron transport throuth the constriction. 
It was demonstrated that the particle flow through the channel with asymmetric constrictions can be controlled not only by changing the size of the constriction but, more strikingly, by tuning the transversal force $f_{y}$ applied to the channel. 
The latter can be easily realized in experiment. 

Our study brings important insights into the dynamics of electrons floating on the surface of superfluid $^{4}$He in channels with constrictions. 
In addition, the present analysis of the charged interacting particles moving through narrow constrictions could be useful for a better understanding of the quasi-one-dimensional and single-file dynamics in other interacting soft-matter or biological systems such as, e.g., colloids or proteins in narrow channels.

\section{Acknowledgments} 

We are thankful to David Rees for providing us with helpful insights in details of the experiments and the corresponding references. 
This work was supported by the Flemish Science Foundation (FWO-Vl) and by the ``Odysseus" program of the Flemish Government and FWO-Vl.


\begin{thebibliography}{50}
\bibitem{voltage} 
We show absolute values of the dimensionless gate voltage; these values can be recalculated to the experimentally measured values by multiplying to a negative constant which can be extracted from fitting the simulation results to the experimental data.
\bibitem{strong} 
The larger value of the driving force is chosen in order to ensure the passing of the particles through the constriction even for moderate values of the transversal force $f_{y}$. 
Otherwise, if we restrict ourselves to very weak driving at $f_{y}=0$ then applying even a very weak $f_{y}>0$ will result in zero transport current throuth the constriction.
\end{thebibliography}
\end{document}